\documentclass[]{pasj02} 
\usepackage[switch,mathlines]{lineno} 

\jyear{2024}
\Received{2024/08/12}
\Accepted{2024/10/01}

\graphicspath{{./}{figures/}} 

\usepackage{comment}
\usepackage{ulem}
\usepackage{url}

\begin{document} 

\title{Self-consistent \textit{N}-body simulation of Planetesimal-Driven Migration\\ \large{I. The trajectories of single planets in the uniform background}}

\author{
 Tenri \textsc{Jinno},\altaffilmark{1}\altemailmark\orcid{0009-0001-5384-4202} \email{223s415s@gsuite.kobe-u.ac.jp} 
 Takayuki R. \textsc{Saitoh},\altaffilmark{1}\altemailmark\orcid{0000-0001-8226-4592}
 \email{saitoh@people.kobe-u.ac.jp} 
 Yoko \textsc{Funato},\altaffilmark{2}\altemailmark\orcid{0000-0002-6992-7600} \email{funato@system.c.u-tokyo.ac.jp}
 and 
 Junichiro \textsc{Makino}\altaffilmark{1}\altemailmark\orcid{0000-0002-0411-4297} \email{jmakino@people.kobe-u.ac.jp}
}
\altaffiltext{1}{Department of Planetology, Graduate School of Science, Kobe University, 1-1 Rokkodai-cho, Nada-ku, Kobe, Hyogo 657-8501, Japan}
\altaffiltext{2}{General Systems Studies, Graduate School of Arts and Sciences, The University of Tokyo, Komaba, Meguro, Tokyo, 153-8902, Japan}




\maketitle

\begin{abstract}
Recent exoplanet observations have revealed a diversity of exoplanetary systems, which suggests the ubiquity of radial planetary migration. One powerful known mechanism of planetary migration is planetesimal-driven migration (PDM), which can let planets undergo significant migration through gravitational scattering with planetesimals. In this series of papers, we present the results of our high-resolution self-consistent $N$-body simulations of PDM, in which gravitational interactions among planetesimals, the gas drag, and Type-I migration are all taken into account. In this first paper (Paper I), we investigate the migration of a single planet through PDM within the framework of the classical standard disk model (the Minimum-Mass Solar Nebula model). Paper I aims to improve our understanding of planetary migration through PDM, addressing previously unexplored aspects of both the gravitational interactions among planetesimals and the interactions with disk gas. Our results show that even small protoplanets can actively migrate through PDM. Such active migration can act as a rapid radial diffusion mechanism for protoplanets and significantly influence the early stages of planetary formation (i.e., during the runaway growth phase). Moreover, a fair fraction of planets migrate outward. This outward migration may offer a potential solution for the ``planet migration problem" caused by Type-I migration and gives a natural mechanism for outward migration assumed in many recent scenarios for the formation of outer planets.
\end{abstract}


\section{Introduction}
Some early planetary formation theories are built upon the classical standard theories  \citep{1972epcf.book.....S,1981PThPS..70...35H,1985prpl.conf.1100H}. In these theories, it is assumed that the terrestrial planets and the
cores of giant planets grow ``in-situ,'' from solid components of the protoplanetary disk around their current locations. Many problems have been pointed out for this in-situ growth model. For instance, it is difficult to explain the formation of Uranus and Neptune within the solar system's lifetime \citep{2001Icar..153..224L,2002AJ....123.2862T}. Protoplanets grown to the size of Mars experience Type-I migration and will drift toward the central star in less than one million years \citep{1986Icar...67..164W,2002ApJ...565.1257T}. Moreover, since the discovery of the first exoplanet by \citet{1995Natur.378..355M}, more than 5,500 exoplanets have been found. A fair fraction of them are hot Jupiters or super-Earths (e.g., \cite{2021ARA&A..59..291Z})\footnote{The data on the current number of discovered exoplanets are available at Exoplanet Exploration Planets Beyond Our Solar System (\url{https://exoplanets.nasa.gov/})}. Their existence cannot be explained by the classic theory of in-situ formation. These discoveries challenge the classical standard theories of planetary formation and necessitate consideration of planetary migration mechanisms. 

In recent years, planetesimal-driven migration (PDM) has gained
attention as a possible mechanism of migration
\citep{2000ApJ...534..428I,2007MsT..........5K,2009Icar..199..197K,2011Icar..211..819C,2014Icar..232..118M,2016ApJ...819...30K}. PDM
is initiated by the emergence of asymmetry in the distribution of
planetesimals around the planets \citep{1984Icar...58..109F}. The
asymmetric distribution leads to an imbalance of the torques acting on the planets, resulting in the self-sustained planetary migration in one direction. The existence of plutinos suggests that Neptune has moved outward slowly during its formation process, and PDM is one possible mechanism for this outward migration \citep{1993Natur.365..819M,1995AJ....110..420M,1999AJ....117.3041H,2003Natur.426..419L}. However, whether or not such an outward migration through PDM actually occurs has been an open question, as realistic simulations in which gravitational interactions among planetesimals (self-stirring), gas drag, and Type-I torque in planetary migration are taken into account have not been conducted
(\cite{2000ApJ...534..428I,2007MsT..........5K,2009Icar..199..197K,2011Icar..211..819C,2014Icar..232..118M,2016ApJ...819...30K}) (see also section \ref{discussion:pdm} and Table \ref{tab2}).

In this series of papers, we investigate the effect of PDM on the planetary formation processes by conducting high-resolution self-consistent $N$-body simulations of PDM, in which gravitational interactions among planetesimals, the aerodynamic gas drag, and Type-I migration are all taken into account. In Paper I, we examine the migration of a single planet through PDM within the classical standard disk model [the Minimum-Mass Solar Nebula model (MMSN)]. Paper I aims to improve our understanding of planetary migration through PDM, specifically by addressing previously unexplored aspects of gravitational interactions among planetesimals and simultaneous interactions with disk gas, including gas drag and Type-I torque. We report the results of 570 self-consistent $N$-body simulations of PDM in Paper I (all models are listed in Table \ref{tab1} in section \ref{method:nemerical_setup}). In the forthcoming papers, starting with Paper II (Jinno et al., in prep), we conduct large-scale $N$-body simulations of planetary formation from a planetesimal disk that extends to the ice giants region (beyond 20 AU), without assuming the presence of a protoplanet as was done in Paper I. Through this series of papers, not only will we clarify the effects of PDM on the planetary formation process, but we will also investigate the entire process of planetary formation, starting from planetesimal disks that range from the regions of terrestrial planets to those of the ice giants.

This paper is organized as follows. In section \ref{method}, we introduce our simulation models and numerical methods. In section \ref{results}, we present the results of 570 self-consistent $N$-body simulations of PDM. In section \ref{discussion}, we briefly summarize the previous studies of PDM and discuss the difference between their simulations and those conducted in this study. Then, we evaluate the PDM criteria presented in \citet{2014Icar..232..118M} by using the results of our simulations. In addition, we discuss the effect of two-body relaxation between planetesimals and protoplanets on planet migration through PDM. In section \ref{summary_conclusion}, we summarize our results and make concluding remarks.

\section{Models and numerical method} \label{method}
\subsection{Disk model} \label{method:disk_model}
We consider a gas disk around a Solar-type star. The gas disk is assumed to be axisymmetric and the gas distribution is similar to the classical model known as the MMSN \citep{1981PThPS..70...35H}. Thus, the gas surface density, the gas density, and the gas disk temperature are given by:
\begin{eqnarray}
\Sigma_{\mathrm{gas}}=&2400f_{\mathrm{gas}}\left(\frac{r}{1\mathrm{AU}}\right)^{p} \exp{\left(-\frac{t}{1\mathrm{Myr}}\right)} ~\mathrm{g~cm}^{-2}, \label{eq:gas_surface_density} \\
\rho_{\mathrm{gas}}=&1.4\times10^{-9}f_{\mathrm{gas}}\left(\frac{r}{1\mathrm{AU}}\right)^{\alpha} \exp{\left(-\frac{t}{1\mathrm{Myr}}\right)}~\mathrm{g~cm}^{-3}, \label{eq:gas_density}\\
T=&2.8\times10^2\left(\frac{r}{1\mathrm{AU}}\right)^{\beta}~\mathrm{K}, \label{eq:temperature}
\end{eqnarray}
where $f_{\mathrm{gas}}$, $r$, and $t$ are the gas scaling factor, the radial distance from the central star, and the time from the beginning of the simulation. Following \citet{1981PThPS..70...35H}, we set values of $f_{\mathrm{gas}}$ to 0.71 and radial dependencies in equations (\ref{eq:gas_surface_density}) to (\ref{eq:temperature}), namely $p$, $\alpha$ and $\beta$ to -3/2, -11/4 and -1/2, respectively. We model the gas dissipation as an exponential decay with a timescale of 1 Myr.

Similar to the gas disk model, the dust distribution is given by:
\begin{equation}
\Sigma_{\mathrm{dust}}=\left\lbrace\begin{array}{ll}
                            40f_{\mathrm{dust}}\left(\frac{r}{1\mathrm{AU}}\right)^{p}~\mathrm{g~cm}^{-2}&(r<r_{\mathrm{ice}}) \\
                            42f_{\mathrm{ice}}\left(\frac{r}{1\mathrm{AU}}\right)^{p}~\mathrm{g~cm}^{-2}&(r\geq r_{\mathrm{ice}}),
                       \end{array} \label{eq:dust_surface_density}
                       \right.
\end{equation}
where $f_{\mathrm{dust}}$ and $r_{\mathrm{ice}}$ are the dust scaling factor and the radial position of the snowline. The values of $f_{\mathrm{dust}}$ and $p$ are set to 0.71 and -3/2. The snowline in our model is assumed to be at 2.0 AU, as it may have been closer to the Sun due to the viscous accretion of the gas disk and the Sun's stellar evolution \citep{2011ApJ...738..141O}.  In addition, $f_{\mathrm{ice}}$ is the scaling factor for the dust beyond the snowline, and it is set to $f_{\mathrm{ice}}=4f_{\mathrm{dust}}=2.84$. The scope of our simulations is restricted to the region beyond the snowline; thus, we use only the latter case of equation (\ref{eq:dust_surface_density}) to represent the dust distribution. 

Here we should note that we set the value of $\Sigma_{\mathrm{dust}}$ to be four times that in the MMSN \citep{2014Icar..232..118M}. Therefore, the gas-to-dust ratio becomes approximately 15, which is four times smaller than that of the MMSN. This ratio is reasonable from the point of view of recent observational studies of protoplanetary disks, such as the Molecules with ALMA at Planet-forming Scales (MAPS) ALMA Large Program (e.g., \cite{2021ApJS..257....1O}), have demonstrated that the gas-to-dust ratio in protoplanetary disks generally ranges from approximately 10 to 100. Thus, the gas-to-dust ratio employed in our simulations is reasonable and well-aligned with the values reported in these studies. However, the gas-to-dust ratio significantly influences the interplay between PDM and Type-I migration. Consequently, we have conducted additional simulations using a disk model with a gas-to-dust ratio based on the MMSN to further investigate the competition between Type-I migration and PDM (see appendix for more details). Our supplementary simulations demonstrate that even when accounting for Type-I migration with a gas-to-dust ratio equivalent to that of the MMSN, protoplanets still exhibit outward migration due to PDM, consistent with the results presented in section \ref{results}.

\subsection{Gas drag model} \label{method:gas_drag_model}
We employ the force formula of gas drag on planetesimals developed by \citet{1976PThPh..56.1756A}:
\begin{equation}
\boldsymbol{F}_{\mathrm{drag}}=-\frac{1}{2}C_{\mathrm{D}}\pi r_{\mathrm{p}}^2\rho_{\mathrm{gas}}|\Delta \boldsymbol{v}|\Delta\boldsymbol{v}, \label{eq:gas_drag}
\end{equation}
where $C_{\mathrm{D}}$, $r_{\mathrm{p}}$ and $\Delta \boldsymbol{v}$ are the gas drag coefficient, the planetesimal radius, and the relative velocity of the planetesimal to the gas disk. According to \citet{1976PThPh..56.1756A}, the gas drag coefficient $C_{\mathrm{D}}$ is approximately equal to unity for the range of planetesimal masses considered in this study (see also Table \ref{tab1}). Thus, we set $C_{\mathrm{D}}=1$ in all of our simulations. Moreover, we assume that planetesimals and protoplanets are perfect spheres. Therefore, the particle radius $r_{\mathrm{p}}$, used in equation (\ref{eq:gas_drag}) is calculated as
\begin{equation}
r_p=\left(\frac{3m_\mathrm{p}}{4\pi \rho_\mathrm{p}}\right)^{1/3}, \label{eq:radius}
\end{equation}
where $m_{\mathrm{p}}$ and $\rho_{\mathrm{p}}$ are the mass and the internal density of the planetesimals. Here we note that we set the value of $\rho_{\mathrm{p}}$ to 2 $\mathrm{g~cm}^{-3}$ in this study.
Furthermore, we assume that the gas disk is laminar and that the magnitude of the disk gas velocity is given by $v_{\mathrm{K}}(1-|\eta|)$ with the local Keplerian velocity $v_{\mathrm{K}}$. Here $\eta$ is a dimensionless quantity that characterizes the pressure gradient of the gas disk which is given by:
\begin{equation}
\eta=-\frac{1}{2}\frac{c_{\mathrm{s}}^2}{r^2\Omega^2}\left(\frac{\partial \log(\rho_{\mathrm{gas}}T)}{\partial \log r} \right), \label{eta}
\end{equation}
where $c_{\mathrm{s}}$ and $\Omega$ are the speed of sound and the Keplerian angular velocity. The functional forms of $c_{\mathrm{s}}$ and $\Omega$ are:
\begin{eqnarray}
c_{\mathrm{s}}&=&\left(\frac{k_{\mathrm{B}}T}{\mu m_{\mathrm{H}}}\right)^{1/2}=1.0\times10^5\left(\frac{T}{280\mathrm~{K}}\right)^{1/2}\mathrm{cm}~\mathrm{s}^{-1},\label{eq:sound}\\
\Omega&=&\left(\frac{GM_{*}}{r^3}\right)^{1/2}=2.0\times10^{-7}\left(\frac{r}{\mathrm{AU}}\right)^{-3/2} ~~\mathrm{s}^{-1},\label{eq:kepler}
\end{eqnarray}
where $k_{\mathrm{B}}$, $\mu=2.34$, $m_{\mathrm{H}}$, $G$ and $M_{*}$ are the Boltzmann constant, the mean molecular weight of the gas, the mass of a hydrogen atom, the gravitational constant and the mass of the central star. 

\subsection{Type-I migration model} \label{method:type-I_model}
For Type-I migration, we follow the model proposed by \citet{2020MNRAS.494.5666I}. It is given by:
\begin{equation}
\frac{\mathrm{d}\boldsymbol{v}}{\mathrm{d}t}=-\frac{v_{\mathrm{K}}}{2\tau_{a}}\boldsymbol{\mathrm{e}}_{\theta}-\frac{v_{r}}{\tau_{e}}\boldsymbol{\mathrm{e}}_{r}-\frac{v_{\theta}-v_{\mathrm{K}}}{\tau_e}\boldsymbol{\mathrm{e}}_{\theta}-\frac{v_z}{\tau_i}\boldsymbol{\mathrm{e}}_{z}, \label{eq:type-I_equation}
\end{equation}
where $\boldsymbol{v}$, $\tau_{a}$, $\tau_{e}$ and $\tau_{i}$ are the planetary velocity, the evolution timescales for the semi-major axis, eccentricity, and inclination. These evolution timescales are given in Appendix D of \citet{2020MNRAS.494.5666I}:
\begin{eqnarray}
\tau_{a}^{-1}&\simeq C_{\mathrm{T}}h^2\left\lbrack1+\frac{C_{\mathrm{T}}}{C_{\mathrm{M}}}(\hat{e}^2+\hat{i}^2)^{1/2}\right\rbrack^{-1} t_{\mathrm{wave}}^{-1}, \label{eq:tau_a}\\
\tau_{e}^{-1}&\simeq0.780\left\lbrack1+\frac{1}{15}(\hat{e}^2+\hat{i}^2)^{3/2}\right\rbrack^{-1}t_{\mathrm{wave}}^{-1}, \label{eq:tau_e}\\
\tau_{i}^{-1}&\simeq0.544\left\lbrack1+\frac{1}{21.5}(\hat{e}^2+\hat{i}^2)^{3/2}\right\rbrack^{-1}t_{\mathrm{wave}}^{-1}, \label{eq:tau_i}
\end{eqnarray}
where $h=H/r$ is the gas disk's aspect ratio, and the scale height of the disk  $H$ is given as:
\begin{equation}
H=\frac{c_{\mathrm{s}}}{\Omega}=5.0\times10^{11}\left(\frac{T_{\mathrm{m}}}{280\mathrm~{K}}\right)^{1/2}\left(\frac{r}{\mathrm{AU}}\right)^{3/2}~~\mathrm{cm}. \label{eq:scaleheight}
\end{equation} 
Additionally, $\hat{e}$ and $\hat{i}$ are the eccentricity and the inclination scaled by $h$. For equation (\ref{eq:tau_a}), $C_{\mathrm{T}}=2.73-1.08p-0.87\beta$, $C_{\mathrm{M}}=-6(2p-\beta-2)$, where $p$ and $\beta$ are given in equations (\ref{eq:gas_surface_density}) and (\ref{eq:temperature}). The characteristic timescale $t_{\mathrm{wave}}$ is given by \citet{2002ApJ...565.1257T} as:
\begin{equation}
t_{\mathrm{wave}}^{-1}=\left(\frac{M_{\mathrm{p}}}{M_{*}}\right)\left(\frac{\Sigma r^2}{M_*}\right)h^{-4}\Omega_{\mathrm{K}}, \label{eq:t_wave}
\end{equation}
where $M_{\mathrm{P}}$ is the planetary mass.

\subsection{Numerical method} \label{method:nemerical_setup}
We conducted a total of 570 self-consistent large-scale $N$-body simulations using GPLUM \citep{2021PASJ...73..660I} on the Japanese supercomputer Fugaku. This section details the $N$-body simulation code GPLUM and the initial conditions of all the simulations.

\subsubsection{$N$-body simulation code GPLUM}\label{method:GPLUM}
GPLUM is an $N$-body simulation code designed for large-scale parallel simulations of planetary formation. This code employs the Framework for Developing Particle Simulator (FDPS), a general-purpose, high-performance library designed for particle simulations \citep{2016PASJ...68...54I,2018PASJ...70...70N}. As a result, GPLUM is noted for its exceptional scalability in parallel computing environments. In the following, we explain the numerical schemes implemented in GPLUM.

GPLUM employs the P$^3$T scheme \citep{2011PASJ...63..881O}, which combines computational techniques of non-collisional and collisional systems to accelerate gravitational calculations between particles. The P$^3$T scheme is a hybrid integrator based on the splitting of the Hamiltonian (e.g., \cite{1999MNRAS.304..793C}), where the system’s Hamiltonian is divided into two parts according to cutoff radii set for each pair of particles. They are called the soft and hard parts and are given by:
\begin{eqnarray}
H&=&H_{\mathrm{Hard}}+H_{\mathrm{Soft}}, \label{Hamiltonian}\\
H_{\mathrm{Hard}}&=&\sum_{i}\left\lbrack\frac{|\boldsymbol{p}_i|^2}{2m_i}-\frac{GM_*m_i}{r_i}\right\rbrack\nonumber\\
&&-\sum_{i}\sum_{j>i}\frac{Gm_im_j}{r_{ij}}[1-W(r_{ij};r_{\mathrm{out},ij})], \label{Hard_part}\\
H_{\mathrm{Soft}}&=&-\sum_{i}\sum_{j>i}\frac{Gm_im_j}{r_{ij}}W(r_{ij};r_{\mathrm{out},ij}), \label{Soft_part}\\
r_{ij}&=&|\boldsymbol{r_{i}}-\boldsymbol{r}_j|, 
\end{eqnarray}
where $m_{i}$,$~\boldsymbol{p}_{i}$, $\boldsymbol{r}_{i}$, $r_{\mathrm{out},ij}$, $W(r_{ij};r_{\mathrm{out},ij})$, and $r_{ij}$ are the mass, momentum, position of the $i$th particle, the cutoff radius, the cutoff function for the Hamiltonian and the distance between $i$th and $j$th particles. The cutoff radius between the $i$th and $j$th particles $r_{\mathrm{out},ij}$ is given by:
\begin{eqnarray}
r_{\mathrm{out},ij}&=&\max(\tilde{R}_{\mathrm{cut},0}r_{\mathrm{Hill},i},\tilde{R}_{\mathrm{cut},1}v_{\mathrm{ran},i}\Delta t,
\tilde{R}_{\mathrm{cut},0}r_{\mathrm{Hill},j},\nonumber\\&&\tilde{R}_{\mathrm{cut},1}v_{\mathrm{ran},j}\Delta t), \label{cut-off_radii}
\end{eqnarray}
where $\tilde{R}_{\mathrm{cut},0}=2$ and $\tilde{R}_{\mathrm{cut},1}=4$ are the non-dimensional parameters, $r_{\mathrm{Hill},i}$ and $r_{\mathrm{Hill},j}$ is the Hill radius, and $v_{\mathrm{ran},i}$ and $v_{\mathrm{ran},j}$ are the root mean square random velocities of particles around the $i$th and $j$th particles, respectively. Here, the random velocity is defined as the difference between the particle’s velocity and the Keplerian velocity and the Hill radius is given by:
\begin{equation}
r_{\mathrm{Hill,i}}=\left(\frac{m_{\mathrm{i}}}{3M_*}\right)^{1/3}a_{\mathrm{i}}, \label{eq:Hill_radius}
\end{equation} where $a_i$ is the semi-major axis of the $i$th particle. As indicated by equation (\ref{cut-off_radii}), GPLUM sets a cutoff radius for each pair of particles. This approach, known as the individual cutoff method, allows for more efficient calculation of particle interactions compared to the shared cutoff method traditionally used in the P$^3$T scheme (e.g., \cite{2011PASJ...63..881O,2017PASJ...69...81I}). The cutoff function $W(r_{ij};r_{\mathrm{out},ij})$ has the following form \citep{2017PASJ...69...81I}:
\begin{equation}
W(y;\gamma)=\left\{
\begin{array}{ll}
\frac{7\gamma^6-9\gamma^5+45\gamma^4-60\gamma^3\ln\gamma-45\gamma^2+9\gamma-1}{3(\gamma-1)^7}y, &\\
    \hspace{4.0cm} (y<\gamma),&\\
    f(y;\gamma)+[1-f(1;\gamma)y], &\\
    \hspace{3.5cm}(\gamma\leq y<1),&\\
        1, &\\
       \hspace{4.0cm} (1\leq y), &\\
\end{array}\right.
\end{equation}
where
\begin{eqnarray}
f(y;\gamma)=&\lbrace-10/3y^7+14(\gamma+1)y^6-21(\gamma^2+3\gamma+1)y^5\nonumber\\
            &+\lbrack 35(\gamma^3+9\gamma^2+9\gamma+1)/3\rbrack y^4\nonumber\\
            &-70(\gamma^3+3\gamma^2+\gamma)y^3\nonumber\\
            &+210(\gamma^3+\gamma^2)y^2-140\gamma^3y\ln y\nonumber\\
            &+(\gamma^7-7\gamma^6+21\gamma^5-35\gamma^4)\rbrace/(\gamma-1)^7.
\end{eqnarray}
The cutoff function indicates that when the distance $r_{ij}$ exceeds their cutoff radius ($r_{ij} > r_{\mathrm{out},ij}$), it becomes unity. Thus, gravitational forces in the hard part only affect particles closer than $r_{\mathrm{out},ij}$.

We can see from equation (\ref{Hard_part}) that the hard part consists of the central star's gravitational potential and the short-range interactions among particles. The time integration of the hard part is performed using the fourth-order Hermite scheme \citep{1991ApJ...369..200M} with the block individual time step scheme \citep{1963MNRAS.126..223A,1986LNP...267..156M}. The soft part given by equation (\ref{Soft_part}) represents the long-range interactions among particles. The soft part is calculated by using the Barnes-Hut tree scheme \citep{1986Natur.324..446B} with constant time step, available in FDPS. Since we use the fourth-order Hermite scheme with the block individual time step scheme for the integration of the hard part, this scheme is not a symplectic method. However, for close encounters where time scales change significantly, it is better to achieve sufficient accuracy with methods that allow for higher order schemes with block individual time steps, rather than with symplectic methods that have difficulty varying time steps. Moreover, typically, the relative energy error is approximately $10^{-5}$ after integrating over $10^5$ years in our simulations. Thus, whether the scheme is symplectic or not is not a problem within the scope of this study.

By employing the P$^3$T scheme with the individual cutoff method, GPLUM enhances the efficiency of splitting gravitational interactions compared to the traditional shared cutoff method, where a single cutoff radius is applied to all particles. In addition, we optimized GPLUM for Fugaku in order to use GPLUM for large-scale $N$-body simulations with over $10^6$ particles distributed in a wide-ranging planetesimal disk that extends beyond 10 AU.

\subsubsection{Initial conditions}\label{method:intial_conditions}
For all simulations, we employ an axisymmetric surface density distribution given by equation (\ref{eq:dust_surface_density}) for the planetesimal disk with inner and outer cutoffs, $r_{\mathrm{in}}$ and $r_{\mathrm{out}}$, as the initial distribution of planetesimals. We set $r_{\mathrm{in}}=2$ AU and $r_{\mathrm{out}}=6$ AU for all models. Initially, eccentricities and inclinations of planetesimals follow a Gaussian distribution with the dispersion $\langle e^2\rangle ^{1/2}=2\langle i^2\rangle ^{1/2}=2r_{\mathrm{Hill}}/a_{\mathrm{p}}$, where $r_{\mathrm{Hill}}$ is the Hill radius of the planetesimal with the semi-major axis $a_{\mathrm{p}}$ \citep{1992Icar...96..107I}. To study the orbital evolution of planets through PDM, we add one planet of the mass 0.1 to 1 $M_{\oplus}$ in a circular orbit of 3 AU. We remove the planetesimals of the same total mass to ensure that the planet is initially placed in the center of a gap in the distribution of planetesimals.

All $N$-body simulations were conducted with a time step for the soft part set at $\Delta t = 2^{-4}~\mathrm{yr}/2\pi \approx 10^{-2}~\mathrm{yr}$. For the hard part, time integration was performed using block individual time steps with the smallest allowed time step set to $\Delta t_{\mathrm{min}} = 2^{-30}~\mathrm{yr}/2\pi \approx 1.5 \times 10^{-10}~\mathrm{yr}$.
Here we note that in all simulations, perfect accretion is assumed for both planetesimal-planetesimal and planet-planetesimal collisions. In other words, the effects of fragmentation were ignored.

The total of 570 simulations are categorized into three groups based on the initial number of particles ($N_{\mathrm{p, init}}$), the number of planetesimals converted into a planet ($M_{\mathrm{p}}/m_{\mathrm{p}}$), the initial planet mass ($M_{\mathrm{p}}$), the simulation end time ($T_{\mathrm{end}}$), the number of runs ($N_{\mathrm{run}}$) and the gas disk-planet interaction\footnote{In this study, the total node hours (NHs) used on Fugaku for the 570 simulations amounted to 9,900 NHs, with the estimated CO$_2$ emissions calculated to be approximately 720 kg-CO$_2$.
\\Sources:\\ Fujitsu Technical Review, March 2020:\\ \url{https://www.fujitsu.com/jp/about/resources/publications/technicalreview/2020-03/article05.html}\\GHG Emissions Accounting, Reporting, and Disclosure System:\\ \url{https://ghg-santeikohyo.env.go.jp/calc}}. Below, we provide an overview of each group (see also Table \ref{tab1}):

\begin{enumerate}
    \item A group dedicated to the investigation of the long-term orbital evolution of planets due to PDM (models 1 and 2).
    \item A group focused on the examination of the mass resolution dependency of PDM (models 3-1 to 3-5).
    \item A group dedicated to the study of the dependence of PDM on the interaction among the gas disk, planetesimals, and planets (models 4-1 to 4-3).
\end{enumerate}

\begin{table*}[h]
\tbl{List of models}{
\begin{tabular}{@{}cccccccccc@{}}
\hline
Group & Model & $N_{\mathrm{p,init}}$  & $m_{\mathrm{p}}$ [$M_{\oplus}$] & $M_{\mathrm{p}}/m_{\mathrm{p}}$ & $M_{\mathrm{p}}$ [$M_{\oplus}$] & $T_{\mathrm{end}}$ [yr] &$N_{\mathrm{run}}$ &Type-I & $F_{\mathrm{drag}}$\\
\hline
1 & 1    & $1.2\times10^5$   & $4.9\times10^{-4}$  & 1035 & 0.5 & $1\times10^{5}$ & 10 & Yes  & Yes\\
1 & 2    & $1.2\times10^5$   & $4.9\times10^{-4}$ & $207\times$ (1--10) & $0.1\times$(1--10) & $1\times10^{5}$ & 10 & Yes  & Yes\\
2 & 3-1    & $1\times10^3$   &  $5.88\times10^{-2}$ & 9 & 0.5 & $5\times10^3$ & 50 & Yes  & Yes\\
2 & 3-2    & $2.5\times10^3$   & $2.35\times10^{-2}$  & 22 & 0.5 & $5\times10^3$ & 50 & Yes  & Yes\\
2 & 3-3    & $5\times10^3$   & $1.17\times10^{-2}$  & 43 & 0.5 & $5\times10^3$ & 50 & Yes  & Yes\\
2 & 3-4    & $1\times10^4$   & $5.88\times10^{-3}$  & 86 & 0.5 & $5\times10^3$ & 50 & Yes  & Yes\\
2 & 3-5    & $3\times10^4$   & $1.96\times10^{-3}$  & 259 & 0.5 & $5\times10^3$ & 50 & Yes  & Yes\\
2 \& 3 & 4-1    & $1.2\times10^5$   & $4.9\times10^{-4}$  & 1035 & 0.5 & $5\times10^3$ & 100 & Yes  & Yes\\
3 & 4-2    & $1.2\times10^5$   & $4.9\times10^{-4}$  & 1035 & 0.5 & $5\times10^3$ & 100 & No & Yes\\
3 & 4-3    & $1.2\times10^5$   & $4.9\times10^{-4}$  & 1035 & 0.5 & $5\times10^3$ & 100 & No & Yes\footnotemark[$*$]\\
\hline
\end{tabular}}\label{tab1}%
\begin{tabnote}
\footnotemark[$*$] The Keplerian gas drag, characterized by the disk gas velocity $v_{\mathrm{K}}$.\\
\end{tabnote}
\end{table*}

All simulations in group 1 were conducted with an initial number of particles $N=1.2\times10^5$ and the simulation time up to $T_{\mathrm{end}}=10^5$ years. Model 1 consists of 10 runs, all with the initial mass of the planet $M_{\mathrm{p}}=0.5~M_{\oplus}$ but from different random number seeds. Model 2 also consists of 10 runs with $M_{\mathrm{p}}=0.1$ to $1.0~M_{\oplus}$.

In group 2, we changed the number of planetesimals from $N=1\times10^3$ to $3\times 10^4$, while keeping the surface mass density the same, and for each value of $N$, we performed 50 runs from different random number seeds. Through these runs, we investigated the effect of the mass ratio between the planet and planetesimals on PDM.

In group 3, we changed the assumptions on Type-I migration and aerodynamic gas drag to study the effect of these model assumptions. Model 4-1 is the standard model which is actually the same as model 1. In model 4-2, we turned off the Type-I migration. In model 4-3, we assume that the orbital velocity of the gas is the same as the Kepler velocity, thereby turning off the drag to the planetesimals in circular orbit. For each model, we made 100 runs.

Table \ref{tab1} shows the group numbers, the names of the models, the initial number of particles ($N_{\mathrm{p,init}}$), the initial mass of a planetesimal
($m_{\mathrm{p}}$), the mass ratio ($M_{\mathrm{p}}/m_{\mathrm{p}}$), the initial mass of a planet ($M_{\mathrm{p}}$), the simulation end time ($T_{\mathrm{end}}$), the number of runs for each model ($N_{\mathrm{run}}$), the presence of Type-I torque and the gas drag ($F_{\mathrm{drag}}$). We note that the magnitude of the disk gas velocity for model 4-3 is set to $v_{\mathrm{K}}$.

\section{Results}\label{results}
\subsection{The planetary migration through PDM} \label{results:long_time}
In this section, we present the results of runs in group 1.

\begin{figure}[hbtp]
 \begin{center}
    \includegraphics[width = 8cm]{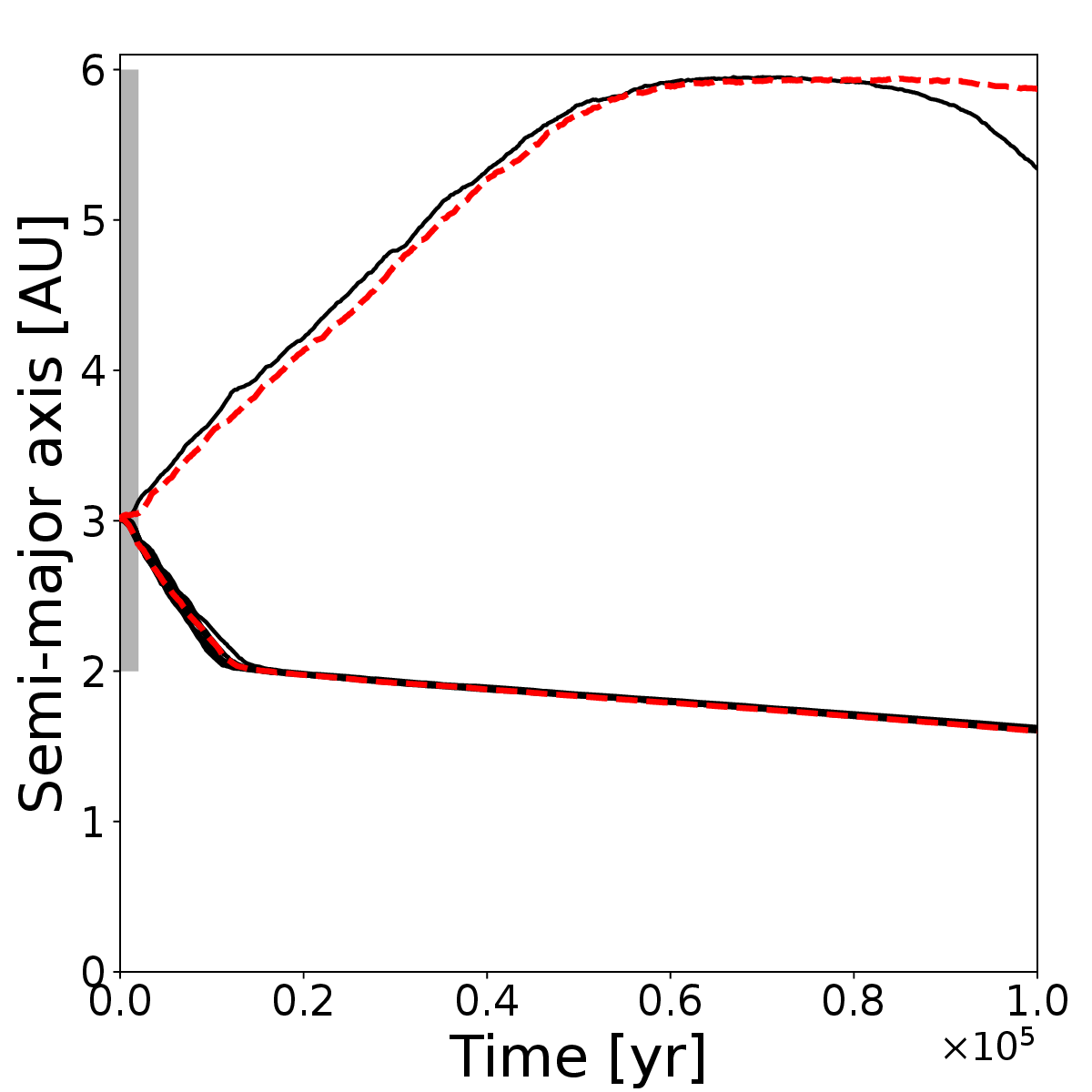}
 \end{center}
 \caption{The time evolution of semi-major axes of protoplanets in model 1. Each line represents the result of 10 runs with different initial seeds. The red dashed lines represent the results of runs shown in figure \ref{fig:snapshots}. The greyed hatch region shows the initial planetesimal disk size. \\{Alt text: The line graph showing the time evolution of semi-major axes of protoplanets in model 1. The horizontal axis represents time, and the vertical axis displays the semi-major axis in AU.} } 
 \label{fig:model1}
\end{figure}

Figure \ref{fig:model1} shows the time evolution of semi-major axes of protoplanets in model 1. In this model, protoplanets exhibited outward PDM in 2 out of 10 runs. Note that each protoplanet was initially placed in a circular orbit, thus we did not give any initial kick to protoplanets. One can observe that once the outward migration started, it continued monotonically up to the outer cutoff of the planetesimal disk at 6 AU. However, when the protoplanet reached the outer cutoff of the planetesimal disk, it reversed direction and began migrating inward. This change in the direction is clearly the effect of the outer cutoff of the disk. In the remaining eight runs, monotonic inward PDM took place and protoplanets reached the inner cutoff of the
planetesimal disk. Here, the change to outward migration was not observed, and protoplanets continued to drift inward through  Type-I migration. In our model, we assumed that gas drag and Type-I torque are active at any radius, even though the initial dust distribution is limited to 2 - 6 AU. 

%

\begin{figure*}[hbtp]
 \begin{center}
    \includegraphics[width = 16cm]{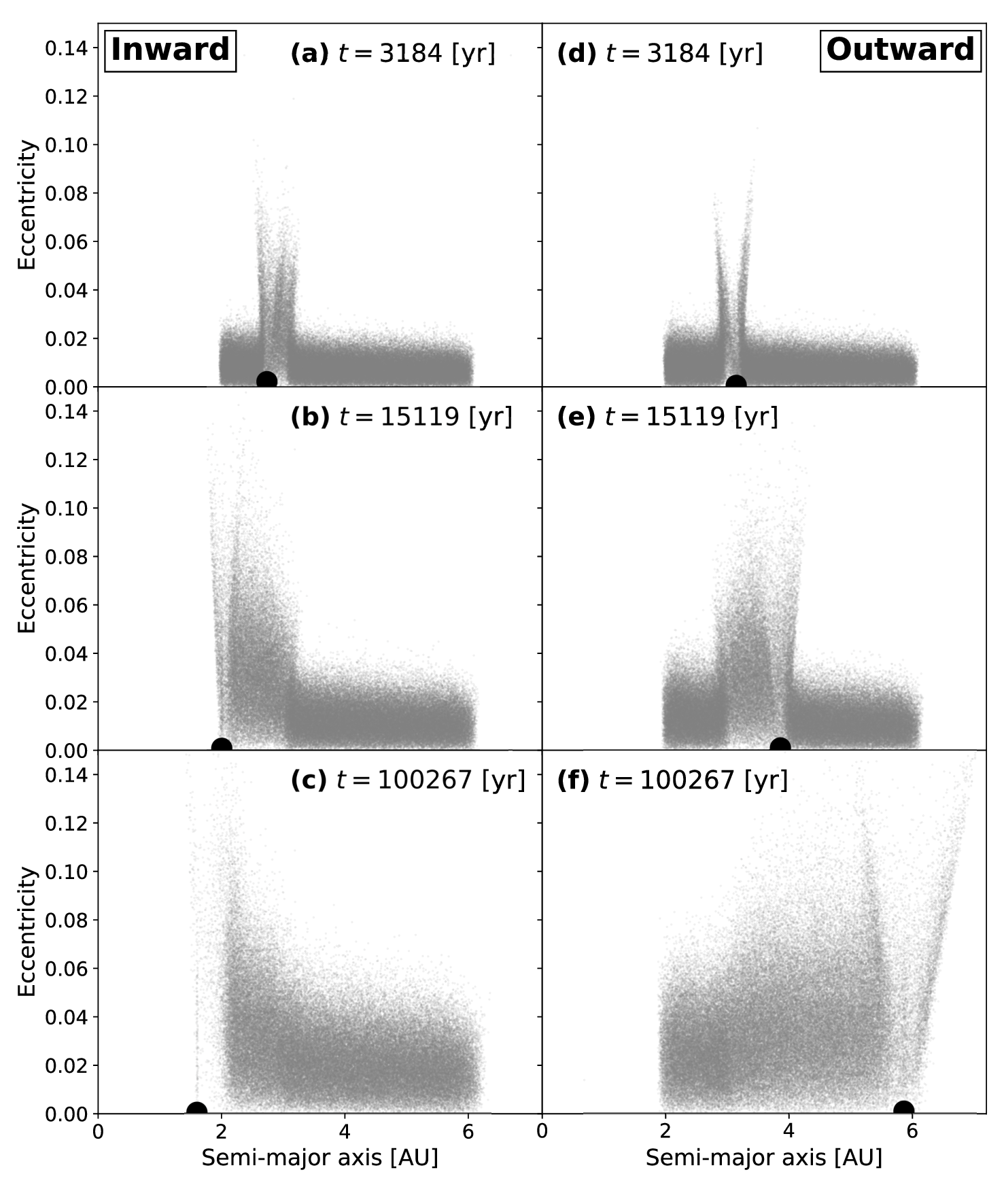}
 \end{center}
 \caption{Snapshots of two simulations from model 1. The vertical and horizontal axes represent eccentricity and semi-major axis. The left (a to c) and right (d to f) sides of the figure show the snapshots of inward PDM and outward PDM runs, arranged chronologically from the top to bottom panel. Black-filled circles and grey dots represent planets and planetesimals, respectively.\\ {Alt text: Snapshots of two simulations from model 1, arranged chronologically from panel (a) to panel (c) (inward PDM) and panel (d) to panel (f) (outward PDM).}} 
 \label{fig:snapshots}
\end{figure*}

 Figure \ref{fig:snapshots} shows the snapshots of two representative runs from model 1 in the eccentricity and semi-major axis plane. In the three panels (a to c) on the left side of figure \ref{fig:snapshots}, the planet migrates inward while gravitationally scattering nearby planetesimals. Conversely, in the three panels (d to f) on the right side, the planet migrates outward. In both inward and outward cases in figure \ref{fig:snapshots}, planetesimals located in the direction of planetary migration are more intensely scattered. This intense interaction between the planet and planetesimals in its migration path leads to an asymmetry in angular momentum transfer among planetesimals on both sides of the planet, thereby inducing monotonic inward/outward PDM. This result indicates that planets can overcome inward Type-I migration and some migrate outward when planet-planetesimal interaction is taken into account.

\begin{figure}[hbtp]
 \begin{center}
    \includegraphics[width= 8cm]{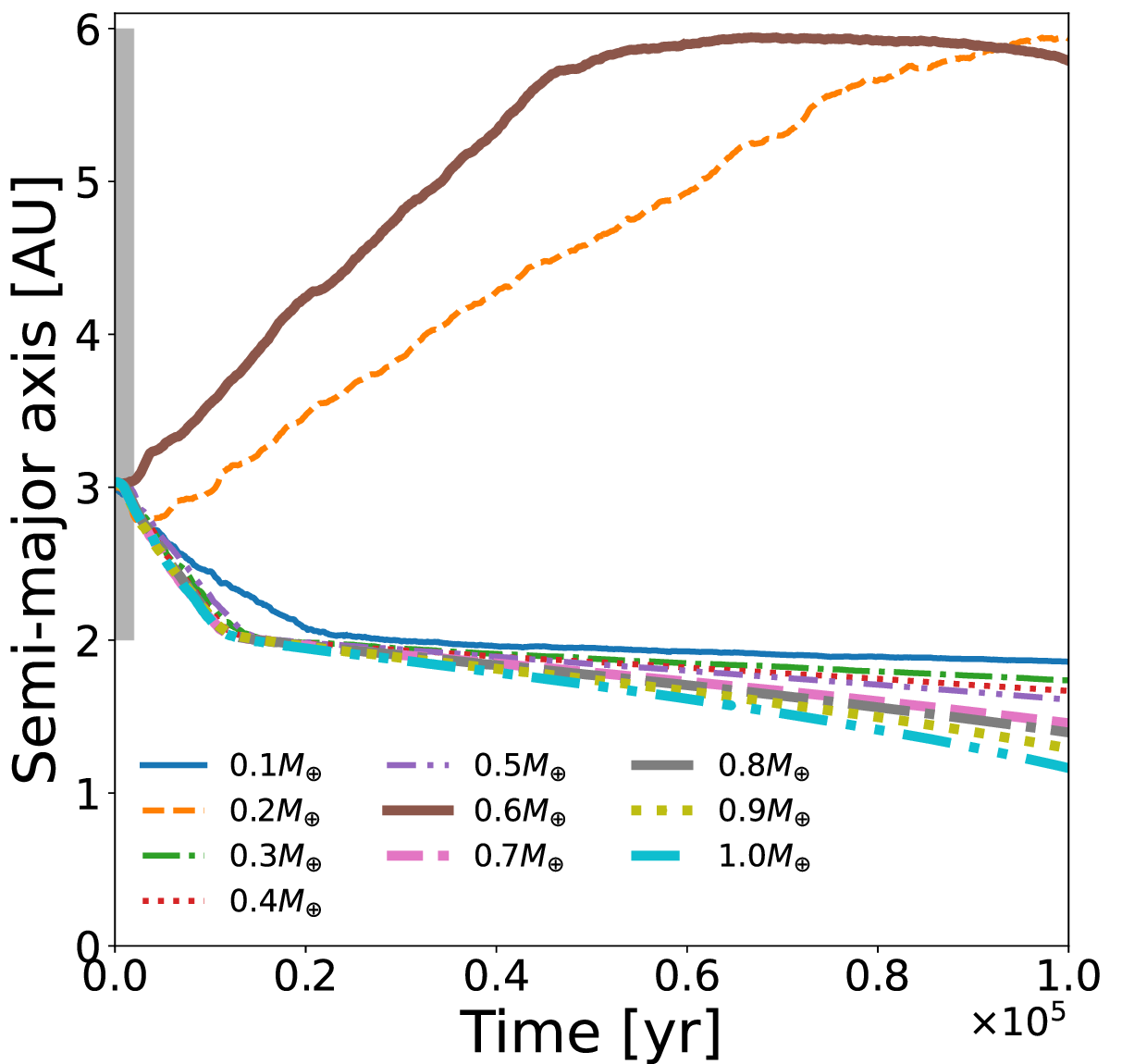}
 \end{center}
 \caption{The time evolution of semi-major axes of protoplanets in model 2. The thin blue solid, orange dashed, green dotted, red dashed-dotted, and purple double-dashed lines represent simulation results for initial planetary masses of 0.1, 0.2, 0.3, 0.4, and 0.5 $M_{\oplus}$, respectively. The thick brown solid, pink dashed, grey dotted, dark-yellow dashed-dotted, and cyan double-dashed lines represent simulation results for initial planetary masses of 0.6, 0.7, 0.8, 0.9, and 1 $M_{\oplus}$, respectively. The greyed hatch region shows the initial planetesimal disk size.\\{Alt text: The line graph showing the time evolution of semi-major axes of protoplanets in model 2. The horizontal axis represents time, and the vertical axis displays the semi-major axis in AU.} } 
 \label{fig:model2}
\end{figure}

Figure \ref{fig:model2} shows the time evolution of semi-major axes of protoplanets in model 2. In this model, runs with initial planet mass $0.2~M_{\oplus}$ and $0.6~M_{\oplus}$ (represented by thin orange dashed and thick brown solid lines, respectively) showed outward PDM. As was the case in model 1, all planets exhibited monotonic PDM both inward and outward except for that in the run with $0.2~M_{\oplus}$. In this particular run, the direction of planet migration through PDM switched from inward to outward around $t=5\times10^3$ years. Another notable feature of this figure is that the migrations of the least massive protoplanets ($0.1~M_{\oplus}$ and $0.2~M_{\oplus}$) seem slower than those of others.

Figures \ref{fig:model1} to \ref{fig:model2} indicate that planets can actively migrate within the protoplanetary disk through PDM both inward and outward even in the presence of Type-I torque.

\subsection{The effect of mass resolution}\label{results:mass_resolution}
It has been believed that whether PDM takes place or not depends on the mass ratio between planetesimals and the
protoplanet \citep{2009Icar..199..197K,2014Icar..232..118M}. Here we investigate this dependency of PDM to the mass ratio using results of $N$-body simulations in group 2 with the number of particles ranging from $10^3$ to $1.2\times10^5$, or mass ratio between the protoplanet and planetesimals 9 to 1035.

\begin{figure*}[hbtp]
  \begin{center}
    \includegraphics[width= 16cm]{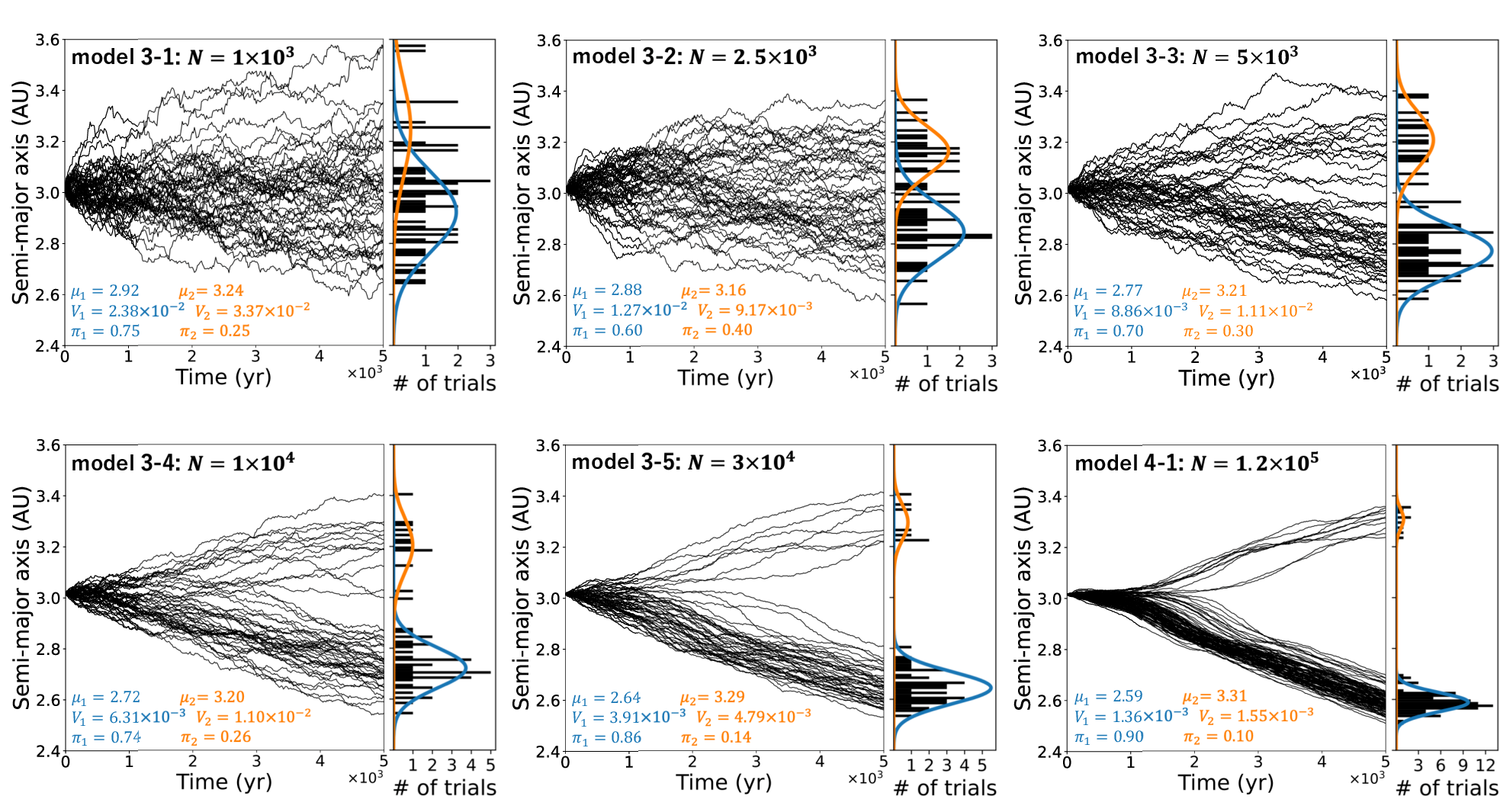}
  \end{center}
 \caption{The time evolution of the semi-major axes of protoplanets and the final distribution of them in models 3-1 to 3-5 and 4-1. Each panel includes 50 runs with different initial seeds with the same number of particles except model 4-1 which includes 100 runs. The initial mass of protoplanets is set to $0.5M_{\oplus}$ in all runs. Histograms are divided into bins of 0.01 AU from 2.4 to 3.6 AU. The two Gaussian functions within each panel are approximations of the histogram using a mixture Gaussian distribution with 2 clusters: $f_{\mathrm{pdf}}(x)=\sum_{i=1}^{2}\pi_{i}/(\sqrt{2\pi} \sigma_{i})\exp{\left[-(x-\mu_{i})^{2}/\sigma_{i}^2\right]}$. Here $x$, $\mu_{i}$, $\sigma_{i}$, and $\pi_{i}$ are the semi-major axis, the expected value, the standard deviation, and the weight of the Gaussian function. We note that when the index $i$ is 1, it represents the parameters of the Gaussian distribution for inward PDM, and when it is 2, it represents the parameters for outward PDM.\\{Alt text: The line graphs and histograms showing the time evolution of the semi-major axes of protoplanets and the final distribution of them. The figure contains 6 subfigures labeled model 3-1 to 3-5 and 4-1.}} 
 \label{fig:trial3_1}
\end{figure*}

Figure \ref{fig:trial3_1} shows the time evolution of the semi-major axes of protoplanets and their final distributions for runs in group 2. From this figure we can see that, as the number of planetesimals increases, planetary migration becomes more monotonic. In other words, for a smaller number of planetesimals, planetary migration is more stochastic. This stochastic planetary migration is different from the diffusion caused by stochastic forces as studied by \citet{2009A&A...497..595R}. Instead, there are two distinct migration trends -both inward and outward- evidenced by the double peak in the final location distribution of the protoplanets. This trend highlights an important but previously not well-investigated nature of PDM.

Note that our result clearly indicates that PDM can drive the migration of protoplanets even when the mass ratio between the
protoplanet and planetesimals is less than 10. Our result is different from the claim of \citet{2014Icar..232..118M} that the mass ratio should be more than 100 for PDM to be effective. They argued that the condition for PDM to be effective is that it is monotonic. They judged if PDM is monotonic or not from a small number of runs with a limited duration in time. Thus, they found a critical mass ratio for monotonicity. Our result implies that the migration can be interpreted as a combined effect of monotonic PDM and stochastic change in the direction caused by the two-body effect, and the latter is larger for a smaller mass ratio. Thus, there is no single critical mass ratio below which PDM is ineffective.

\begin{figure}[hbtp]
  \begin{center}
    \includegraphics[width= 8.0cm]{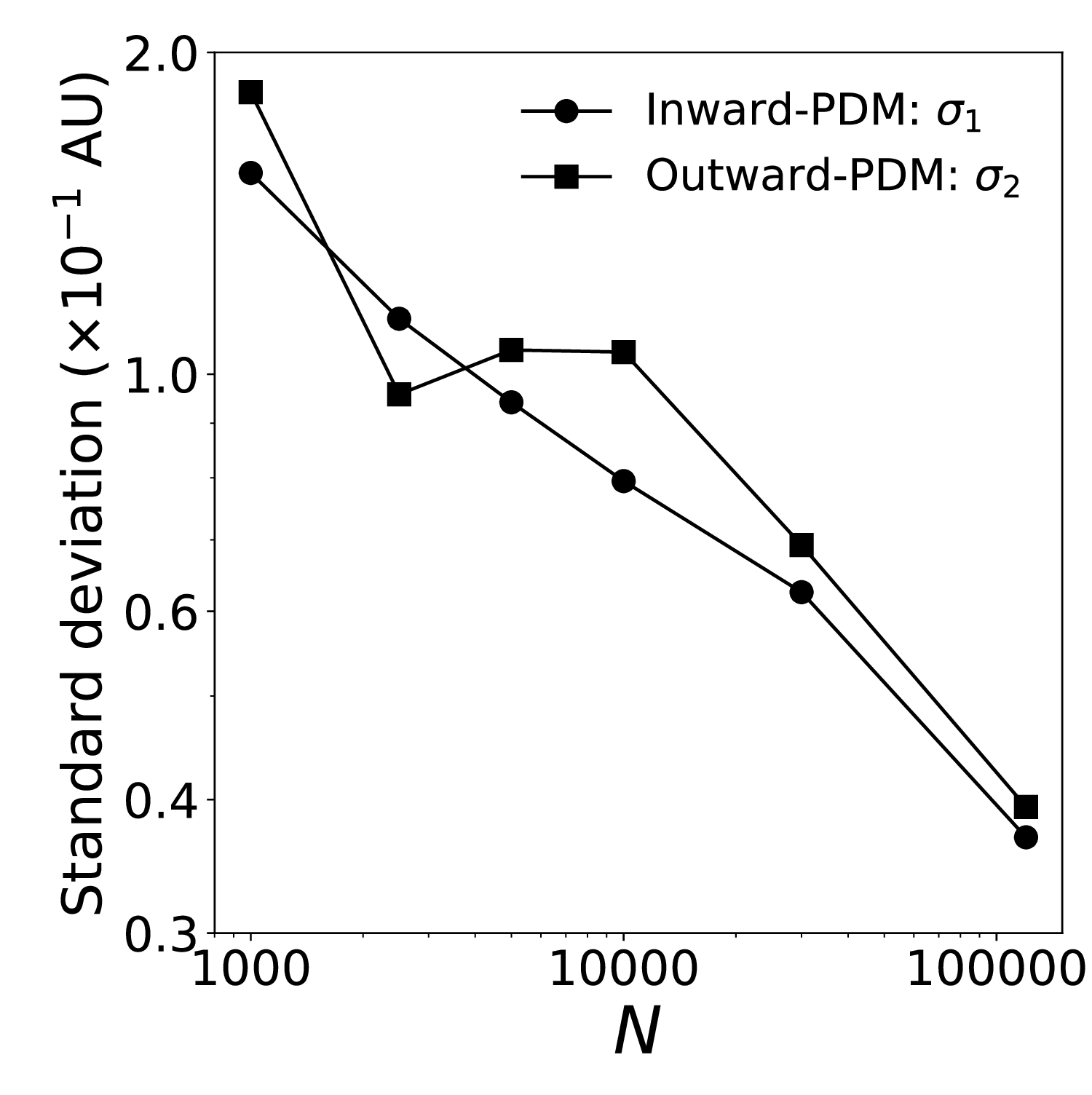}
   \end{center}
\caption{The standard deviation  $\sigma_1$ and $\sigma_2$ of the final location of protoplanets plotted as functions of the number of planetesimals $N$. Circles and squares represent inward and outward PDM, respectively.\\{Alt text: The line graph with vertical and horizontal axes showing the standard deviation of the final positions of the protoplanets for models 3-1 to 4-1 and the number of planetesimals respectively.}} 
 \label{fig:dispersion}
\end{figure}

\begin{figure}[hbtp]
 \begin{center}
    \includegraphics[width=8cm]{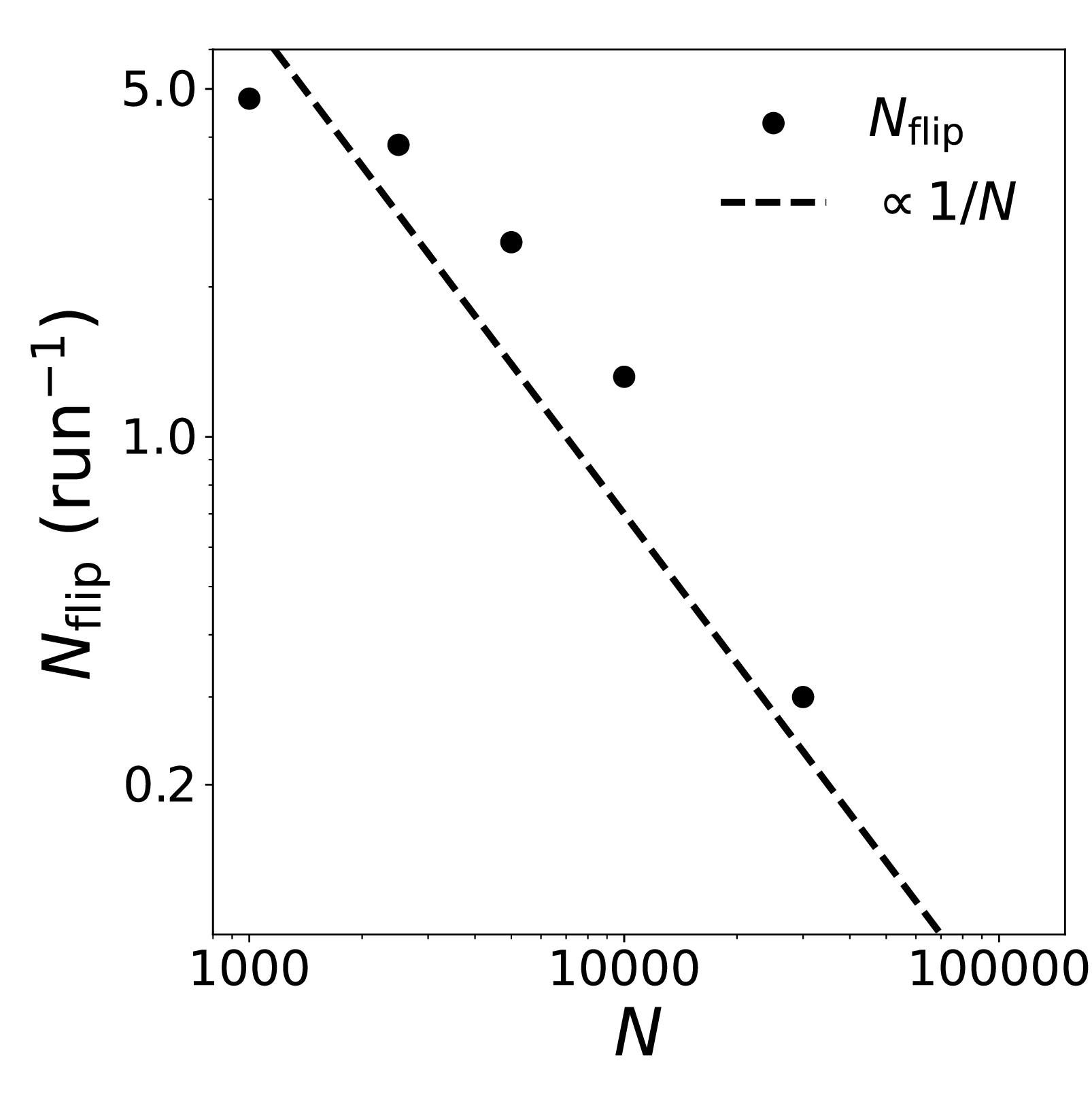} 
 \end{center}
 \caption{The average flip frequency of planetary migration direction per run for models 3-1 to 3-5. Given that a ``flip" means a change in the planetary migration direction, we define a flip as the point in time when the time derivative of the planet's semi-major axis becomes zero (see section \ref{discussion:perturbation}). We use the Savitzky-Golay (SG) filter to smooth the time evolution of the planetary semi-major axis to investigate global changes in planetary migration direction \citep{1964AnaCh..36.1627S}.\\{Alt text: The scatter plot with vertical and horizontal axes showing the averaged flip frequency of planetary migration direction per run for models 3-1 to 3-5 and the number of planetesimals respectively.}} 
 \label{fig:flipping}
\end{figure}

Figure \ref{fig:dispersion} shows the standard deviation of the final location of protoplanets as a function of $N$. We can see that the deviation is roughly proportional to $1/\sqrt{N}$. This trend can be understood by looking at the frequency of directional flips in planetary migration $N_{\mathrm{flip}}$ since a higher flip frequency should imply that the PDM is more stochastic.  Figure \ref{fig:flipping} shows $N_{\mathrm{flip}}$ as a 
function of $N$. This figure indicates that $N_{\mathrm{flip}}$ is proportional to $N^{-1}$. Thus, in models with fewer particles, the increase in the flip frequency causes inhibition of the monotonous planetary migration through PDM. As a result, the distribution of the final locations becomes wider. This $N^{-1}$ dependence can be understood as the effect of two-body encounters between planetesimals and protoplanets
\citep{2008gady.book.....B} (see also section \ref{discussion:perturbation}). 

\begin{figure}[hbtp]
 \begin{center}
    \includegraphics[width=8cm]{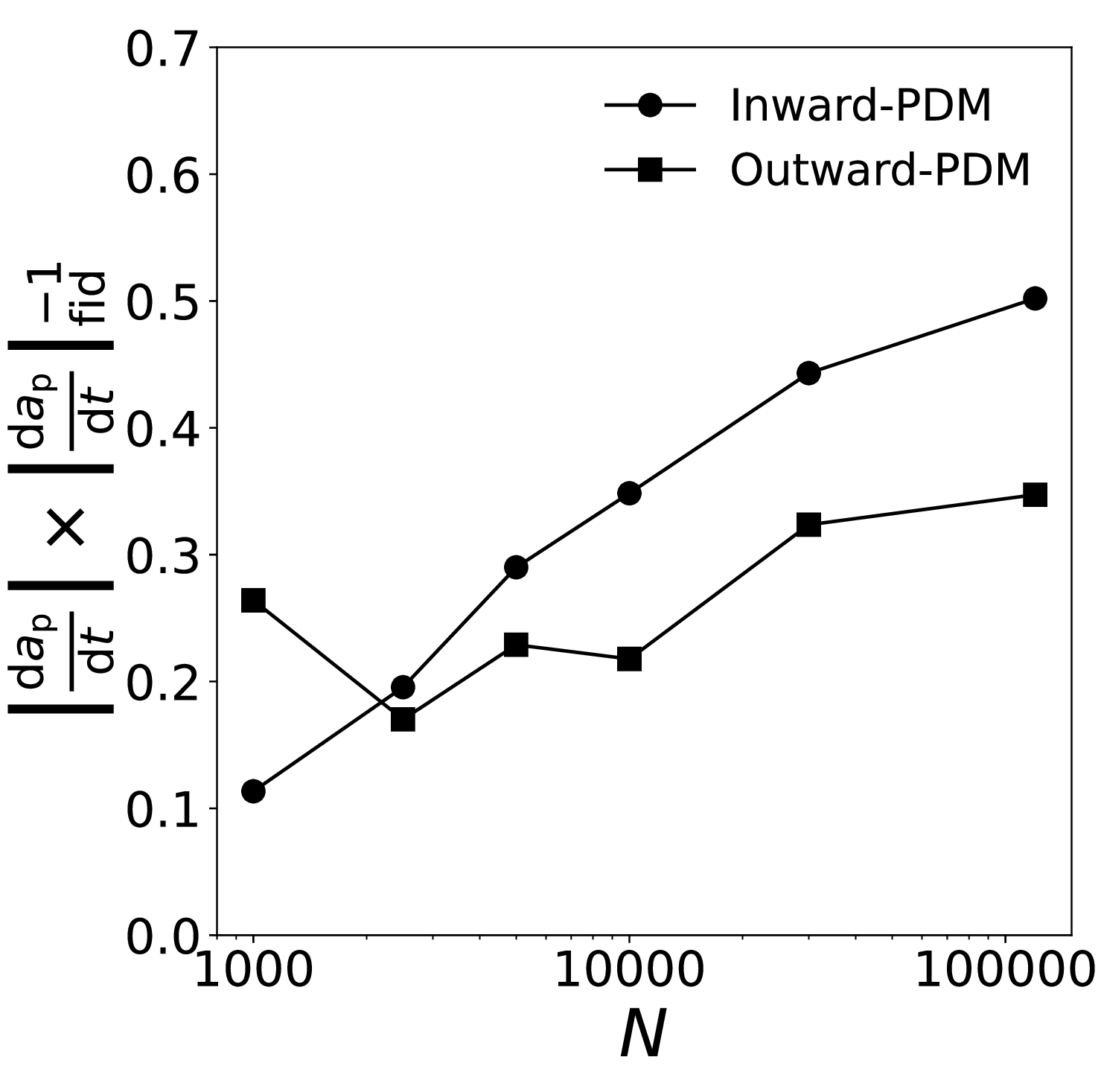}
 \end{center}
 \caption{The normalized average migration rate of models 3-1 to 3-5 and 4-1. Circles and squares represent inward and outward PDM, respectively.\\{Alt text: The line graph, with vertical and horizontal axes, showing the normalized averaged migration rate from models 3-1 to 4-1 and the number of planetesimals respectively.}} 
 \label{fig:fiducial}
\end{figure}

Figure \ref{fig:fiducial} shows the average migration rates in group 2. The migration rates are normalized by the fiducial migration rate given by
\citep{2000ApJ...534..428I,2009Icar..199..197K},
\begin{equation}
\left|\frac{\mathrm{d}a_{\mathrm{p}}}{\mathrm{d}t}\right|_{\mathrm{fid}}\approx \frac{4\pi\Sigma_{\mathrm{dust}}a_{\mathrm{p}}^2}{M_{\mathrm{Sun}}}\frac{a_{\mathrm{p}}}{T}, \label{eq:fiducial}
\end{equation}
where $a_{\mathrm{p}},~\Sigma_{\mathrm{dust}}$ and $T$ are the planet's semi-major axis, the surface density of the planetesimal disk, and the orbital period of the planet. It should be noted that the averaged migration rate is calculated by dividing the expected value of the protoplanets's final distribution, $\mu_{i}~(i=1,2)$, shown in figure \ref{fig:trial3_1}, by the simulation duration time of $5\times10^3$ years. Here we also note that in determining the specific values for the fiducial migration rate given in equation (\ref{eq:fiducial}), we use the dust surface density distribution and orbital period at 3 AU. Figure \ref{fig:fiducial} shows that the normalized migration rate is higher for large $N$ and reaches 0.5 for inward migration and 0.35 for outward migration. Our result is consistent with that of \citet{2009Icar..199..197K}, who suggested that the normalized  PDM rates were typically in the range of  $0.2-0.7$  and that the inward migration rate is higher than the outward migration rate. However, we have clearly shown that PDM rate only gradually decreases when we reduce the number of planetesimals (and thus the mass ratio).

Figure \ref{fig:flipping} shows that the monotonic nature of PDM sufficiently converges in model 4-1 ($N=1.2\times10^5$). This indicates that the monotonic nature will also appear in larger $N$ cases, even in cases of $N\sim10^8$, i.e., the realistic $N$ cases\footnote{In model 4-1 ($N=1.2\times10^5$), the planetesimal radius is about 700 km (refer to table \ref{tab1} and equation (\ref{eq:radius})). However according to \citet{2009Icar..204..558M}, typical planetesimal radii are considered to be around 50 km. To realize this size in simulations requires a radius reduction by an order of magnitude. Thus, we need to increase $N$ by three orders of magnitude to around $N\sim10^8$.}. In figure \ref{fig:dispersion}, although it has not yet converged within the range of particle numbers used in our study ($N=10^3 - 1.2\times10^5$), the standard deviation of the final locations of protoplanets is proportional to $1/\sqrt{N}$. It can be expected, therefore, that a further increase in the number of particles will result in a narrower distribution of the final locations of protoplanets. The standard deviation will be $\sim 0.0006$ AU for $N\sim 10^8$, which is much smaller than the interplanetary distance. From the point of view of planetary formation, it is practically equivalent to 0. In figure \ref{fig:fiducial}, although the average migration rate has not converged, the slopes decrease as $N$ increases. This suggests that the average migration rate might converge with further increases in $N$. The summary of what figures \ref{fig:dispersion} and \ref{fig:fiducial} show is as follows. It can be expected that the final location of protoplanets will converge if the number of particles is further increased.

\subsection{The effect of gas disk-planet interaction} \label{results:disk_planet}
In the presence of a gas disk, planets undergo Type-I migration, and gas drag influences the dispersion of planetesimal velocities. These interactions may affect the way PDM actually works \citep{2011Icar..211..819C,2016ApJ...819...30K}. Here, we present the results of our $N$-body simulations in group 3 with and without Type-I torque and in Keplerian gas environments to examine the influence of the gas disk on PDM.

\begin{figure*}[hbtp]
 \begin{center}
      \includegraphics[width= 16cm]{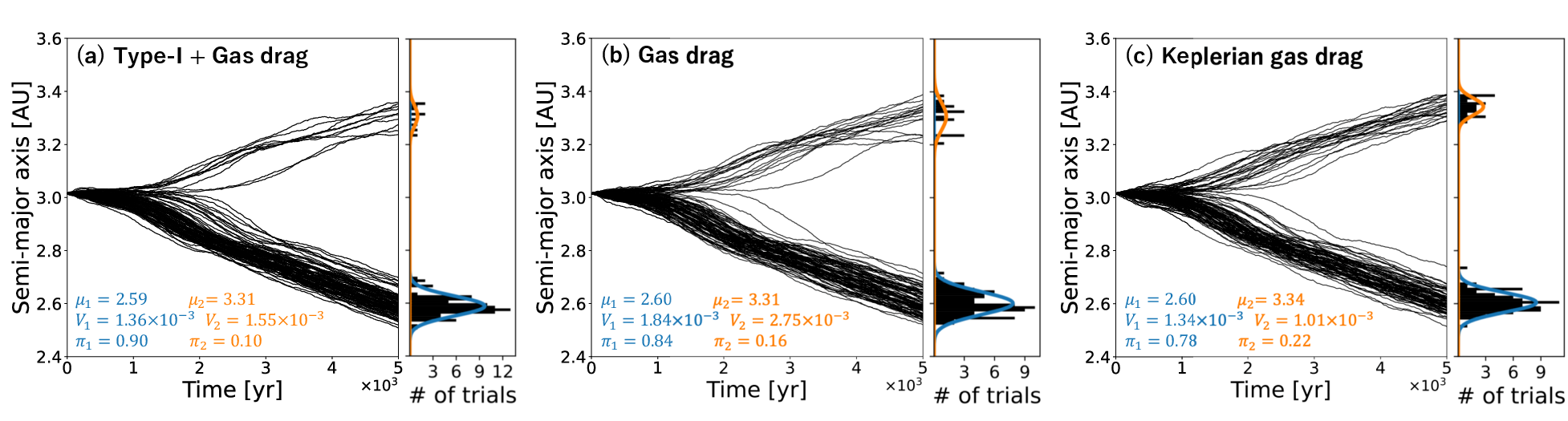}
 \end{center}
 \caption{The time evolution of the semi-major axes of protoplanets and the final distribution of them in models 4-1 to 4-3. The left to the right panels represent models with (a) Type-I torque and the gas drag, (b) the gas drag, and (c) the Keplerian gas drag, respectively. The initial mass of protoplanets is set to $0.5M_{\oplus}$ in all runs. Each histogram is divided into bins of 0.01 AU from 2.5 to 3.5 AU. Same as figure \ref{fig:trial3_1}, the two Gaussian functions within each panel represent approximations of the histograms using a mixture Gaussian distribution with 2 clusters. The legends in each panel represent the parameter for mixture Gaussian distributions of models 4-1 to 4-3. These parameters are the expected values $\mu_{i}$, the variances $V_{i}$, and weights $\pi_{i}$ of the Gaussian distributions. When the index $i$ is 1, it represents the parameters of the Gaussian distribution for inward PDM, and when it is 2, it represents the parameters for outward PDM. It should be noted that the parameters $\pi_1$ and $\pi_2$ correspond to the fraction of samples in each respective cluster.\\{Alt text: Line plots and histograms showing the time evolution of the semi-major axes of protoplanets and their final distribution. The figure contains 3 subplots labeled (a) Type I plus Gas drag, (b) Gas drag, and (c) Keplerian gas drag.}} 
 \label{fig:trial4_1}
\end{figure*}

Figure \ref{fig:trial4_1} shows the time evolution of the semi-major axes of protoplanets and their final distributions for runs in group 3. The fraction of protoplanets that exhibit outward migration is the lowest for model 4-1 (both gas drag and Type-I are on), higher for model 4-2 (Type-I is off), and even higher for model 4-3 (Type-I is off and the gas is assumed to have Kepler velocity). Thus, in our setting the existence of gas drag and Type-I torque do reduce the probability of outward migration, but even when they are included outward migration does occur.

\section{Discussion}\label{discussion}
\subsection{Previous studies of PDM}\label{discussion:pdm}
The effect of PDM on planetary migration has been investigated in several previous studies \citep{2000ApJ...534..428I,2007MsT..........5K,2009Icar..199..197K,2011Icar..211..819C,2014Icar..232..118M,2016ApJ...819...30K}. In this subsection, we briefly summarize each of these studies in sections \ref{discussion:ida} to \ref{discussion:Kominami} and then discuss the differences between their works and present work in section \ref{discussion:comparison}. We give the summary of the comparison in Table \ref{tab2}.

\subsubsection{\citet{2000ApJ...534..428I}}\label{discussion:ida}
\citet{2000ApJ...534..428I} simulated the orbital evolution of proto-Neptune through PDM. They conducted a Monte Carlo simulation that took into account the gravitational interactions between planetesimals and proto-Neptune, demonstrating that proto-Neptune can undergo outward PDM due to the asymmetry in the planetesimal distribution. Furthermore, they showed that this asymmetry in the planetesimal distribution is self-sustained through scattering and the migration of proto-Neptune. They also calculated the net transfer of angular momentum by planetesimal scattering and derived a fiducial migration rate for PDM (equation (17) in this paper). Their simulation initially placed a proto-Neptune with a mass of $M_{\mathrm{p}} = 3 \times 10^{28}$ g ($\sim 0.3 M_{\mathrm{Nept}}$) at 25 AU, distributing planetesimals only from 25 AU to 25 AU + 25 $r_{\mathrm{Hill}}$ and omitting the inner region. Moreover, they ignored the gravitational interactions among planetesimals (self-stirring) and assumed a gas-free environment, thus neglecting effects such as gas drag and Type-I torques.

\subsubsection{\citet{2007MsT..........5K} \& \citet{2009Icar..199..197K}}\label{discussion:kirsh}
Unlike \citet{2000ApJ...534..428I}, these two studies have set up planetesimal disks both inside and outside of the protoplanet and investigated planet migration through PDM using $N$-body simulations. A key aspect of their research is the use of the $N$-body simulation code SyMBA \citep{1998AJ....116.2067D} to examine planet migration through PDM across a broad parameter space, including planet mass, disk eccentricity, horseshoe mass, the initial semi-major axis of the protoplanet, and disk surface density distribution. Their simulations showed that planets migrate more significantly through PDM when the cold disk is assumed (i.e., where planetesimals have lower random velocities), and the mass of planetesimals in the closed encounter zone (2.5 $r_{\mathrm{Hill}}$ width around the protoplanet) is greater than that of the planet. Furthermore, their results demonstrate that outward PDM results in slower migration speeds compared to inward PDM. This outcome is explained by the fact that planetesimals located within the inner regions of the disk have shorter encounter periods and a higher surface density distribution, leading to stronger scattering of these inner planetesimals by the protoplanet. While these studies provide a detailed examination of PDM trends across a wide parameter space, their use of SyMBA means that gas drag and Type-I torques were not considered, and gravitational interactions among planetesimals were also neglected. Importantly, \citet{2009Icar..199..197K} discarded the outcomes of outward PDM, and thus did not present detailed trends of outward PDM.

\subsubsection{\citet{2011Icar..211..819C}}\label{discussion:cappobianco}
A distinctive feature of \citet{2011Icar..211..819C} is their implementation of $N$-body simulations of PDM that consider gas drag and Type-I torques, factors previously neglected in the three studies mentioned earlier. Their simulations revealed that gas drag significantly mitigates planetesimal scattering, thereby substantially influencing planet migration through PDM. They also found that while planets typically undergo inward migration when interacting with sufficiently large or small mono-dispersed planetesimals, outward migration is more common with planetesimals sized between $0.5-5$ km. However, similar to \citet{2007MsT..........5K} \& \citet{2009Icar..199..197K}, their study also used the SyMBA code (albeit incorporating gas drag and Type-I torques), resulting in the neglect of gravitational interactions among planetesimals. Furthermore, due to limitations in mass resolution, each particle in their simulations is assumed to be a tracer of a planetesimal, artificially subjecting them to strong gas drag.

\subsubsection{\citet{2014Icar..232..118M}}\label{discussion:ML14}
\citet{2014Icar..232..118M} investigated the effect of PDM within the terrestrial planet formation region. A significant aspect of their study is the detailed investigation of the conditions driving PDM (for detailed PDM criteria, see section \ref{discussion:pdm_criteria}). Additionally, they employ a planetesimal disk containing multiple protoplanets, constructed using the Monte Carlo method for the initial conditions of their $N$-body simulations. Thus, unlike previous studies that assumed a single planet, they consider the effects of PDM in a more realistic system. However, their $N$-body simulations assume a gas-free environment, thus neither Type-I migration nor gas drag is considered. Additionally, as previous studies mentioned above, they did not account for gravitational interactions among planetesimals.

\subsubsection{\citet{2016ApJ...819...30K}}\label{discussion:Kominami}
\citet{2016ApJ...819...30K} is notable for using the $N$-body simulation code Kninja, which is optimized for the K computer, to conduct the world’s first $N$-body simulations of PDM that account for gravitational interactions among planetesimals. Their simulations reveal that viscous stirring caused by the gravitational interactions among planetesimals could heat the entire disk, potentially suppressing planet migration through PDM. However, they assume a gas-free environment, thus effects such as gas drag and Type-I torque are not considered. Additionally, due to the high computational costs of the gravitational interactions among planetesimals, only 4 simulations were performed, limiting the detailed investigation of planet migration driven by PDM. 

\subsubsection{Comparison with previous studies}\label{discussion:comparison}
As summarized in sections \ref{discussion:ida} through \ref{discussion:Kominami} and Table \ref{tab2}, all previous studies except for \citet{2016ApJ...819...30K} have neglected gravitational interactions among planetesimals, thereby naturally neglecting the effect of viscous stirring in their simulations. Viscous stirring plays a crucial role on PDM as it increases the random velocities of planetesimals. This increase subsequently reduces the frequency of their close encounters with a planet, thereby suppressing planetary migration through PDM. Our study, in contrast, takes into account gravitational interactions among planetesimals in all simulations, providing a detailed examination of how viscous stirring influences planetary migration through PDM. Furthermore, while \citet{2016ApJ...819...30K} did consider the gravitational interactions among planetesimals, they only conducted 4 simulations due to the high computational cost, leaving the trend in PDM unexplored. On the other hand, we conducted 570 simulations across a wide parameter space that incorporate gravitational interactions among planetesimals, enabling a comprehensive statistical analysis of the trends in planetary migration through PDM.

Previous studies, with the exception of \citet{2011Icar..211..819C}, have assumed a gas-free environment, thereby neglecting the effects of gas drag and Type-I migration. Gas drag acts to suppress the increase in random velocities of planetesimals, thereby facilitating planetary migration through PDM. Consequently, gas drag, along with viscous stirring, is crucial for understanding the effects of PDM on planetary migration. Furthermore, since Type-I migration induces radial migration of planets, it is also important to examine its interaction with PDM. Thus, in this study, we considered a gas disk in our simulations to investigate the effects of gas drag and Type-I migration on planetary migration through PDM. Here, we should note that while \citet{2011Icar..211..819C} claims to consider these effects, their methodology involved artificially strong application of gas drag to planetesimals which we have consciously avoided. Instead, we employed realistic gas drag based on the MMSN (see section \ref{method:gas_drag_model}).

Although our simulations consider gravitational interactions among planetesimals, gas drag, and Type-I migration, which are neglected in \citet{2014Icar..232..118M}, they are still idealized compared to the multi-protoplanet systems studied by \citet{2014Icar..232..118M}. In the following papers, however, this limitation will be resolved by conducting large-scale, self-consistent $N$-body simulations from a planetesimal disk, without assuming the presence of protoplanets.

\begin{table}[h]
\tbl{Comparison with previous studies}{
\begin{tabular}{@{}lccc@{}}
\hline
Authors & Self-stirring & Gas drag & Type-I\\
\hline
\citet{2000ApJ...534..428I}    & No  & No  & No\\
\citet{2007MsT..........5K}& No  & No  & No\\
\citet{2009Icar..199..197K}    & No  & No  & No\\
\citet{2011Icar..211..819C}    & No  & Yes\footnotemark[$*$]  & Yes\\
\citet{2014Icar..232..118M}    & No  & No  & No\\
\citet{2016ApJ...819...30K} & Yes  & No  & No\\
Jinno et al. (This paper)    & Yes  & Yes  & Yes\\
\hline
\end{tabular}}\label{tab2}%
\begin{tabnote}
\footnotemark[$*$] Artificially strong gas drag to counterbalance mass resolution limitations.\\
\end{tabnote}
\end{table}

\subsection{The PDM criteria}\label{discussion:pdm_criteria}
\citet{2014Icar..232..118M} identified five criteria that have to be satisfied simultaneously in order for planets to migrate due to PDM. Here we discuss the validity of these criteria taking into account the result of our simulations. 

The five criteria are as follows:
\begin{enumerate}
\item Mass ratio criterion:\\
The amount of planetesimals within $5 r_{\mathrm{Hill}}$ of the planet at $a_{\mathrm{p}}$ need to be larger than $\frac{1}{3}M_{\mathrm{p}}$.
\item Mass resolution criterion:\\
The mass ratio between the planet and the planetesimal should be larger than 100.
\item Disk eccentricity criterion:\\
The eccentricity of the planetesimal disk surrounding the planet has to be less than $5 \times$ the reduced hill ($ r_{\mathrm{Hill}}/a_{\mathrm{p}}$).
\item Crowded criterion:\\
If there are two or more planets near each other in the disk, the larger and the smaller planets' mass ratio needs to be greater than 10, for the larger planet to migrate by PDM.
\item Growth timescale criterion:\\
The planet's growth timescale, $\tau_{\mathrm{grow}}$, needs to be longer than the planet's migration timescale, $\tau_{\mathrm{mig}}$. This is related to criterion 4, which implies that the bodies surrounding the planet should not become larger than the planet itself.
\end{enumerate}
\begin{figure}[hbtp]
 \includegraphics[width= 8cm]{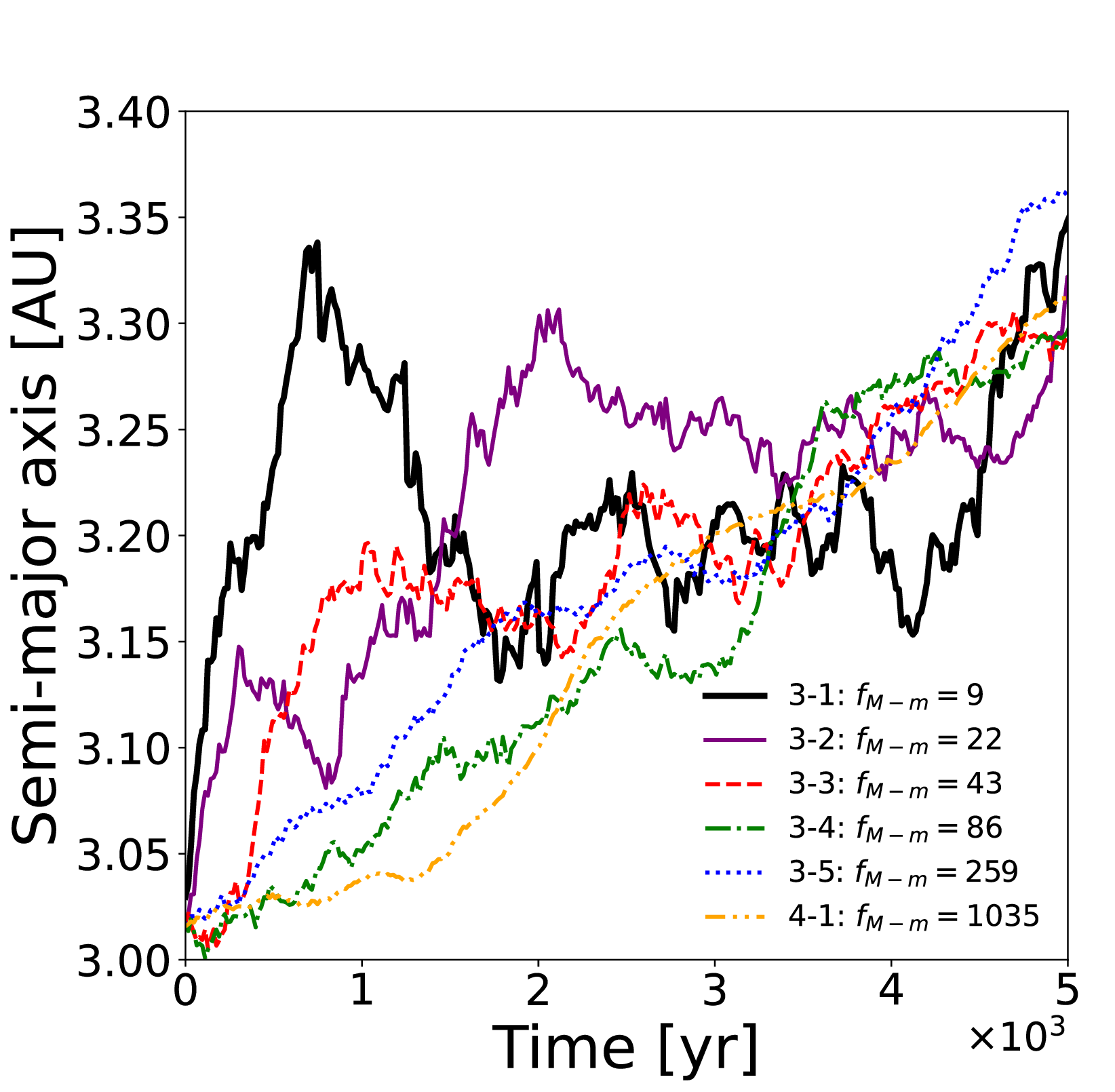}
 \caption{The time evolution of the semi-major axes of protoplanets in models 3-1 to 3-5 and 4-1. The thick solid line (black), thin solid line (purple), dashed line (red), dash-dotted line (green), dotted line (blue), and double-dotted dashed line (orange) represent the results from models 3-1, 3-2, 3-3, 3-4, 3-5, and 4-1, respectively.\\{Alt text: The line graph showing the time evolution of semi-major axes of protoplanets in models 3-1 to 3-5 and 4-1. The horizontal axis represents time in thousands of years, and the vertical axis shows the semi-major axis in AU.}
 \label{fig:criteria_0}} 
\end{figure}

In our study, we examine the trajectory of a single planet in a uniform background. Thus, we check the first three criteria only, since the last two are related to the cases where there are multiple planets in the disk. Figure \ref{fig:criteria_0} shows the time evolution of planetary orbits in models 3-1 to 3-5 and 4-1. These six trials were selected from one trial in each model where outward PDM occurred. Below, we examine the first three criteria for these six trials.

\subsubsection{The mass ratio criterion}\label{discussion:mass_ration_criterion}
\begin{figure}[hbtp]
 \includegraphics[width= 8cm]{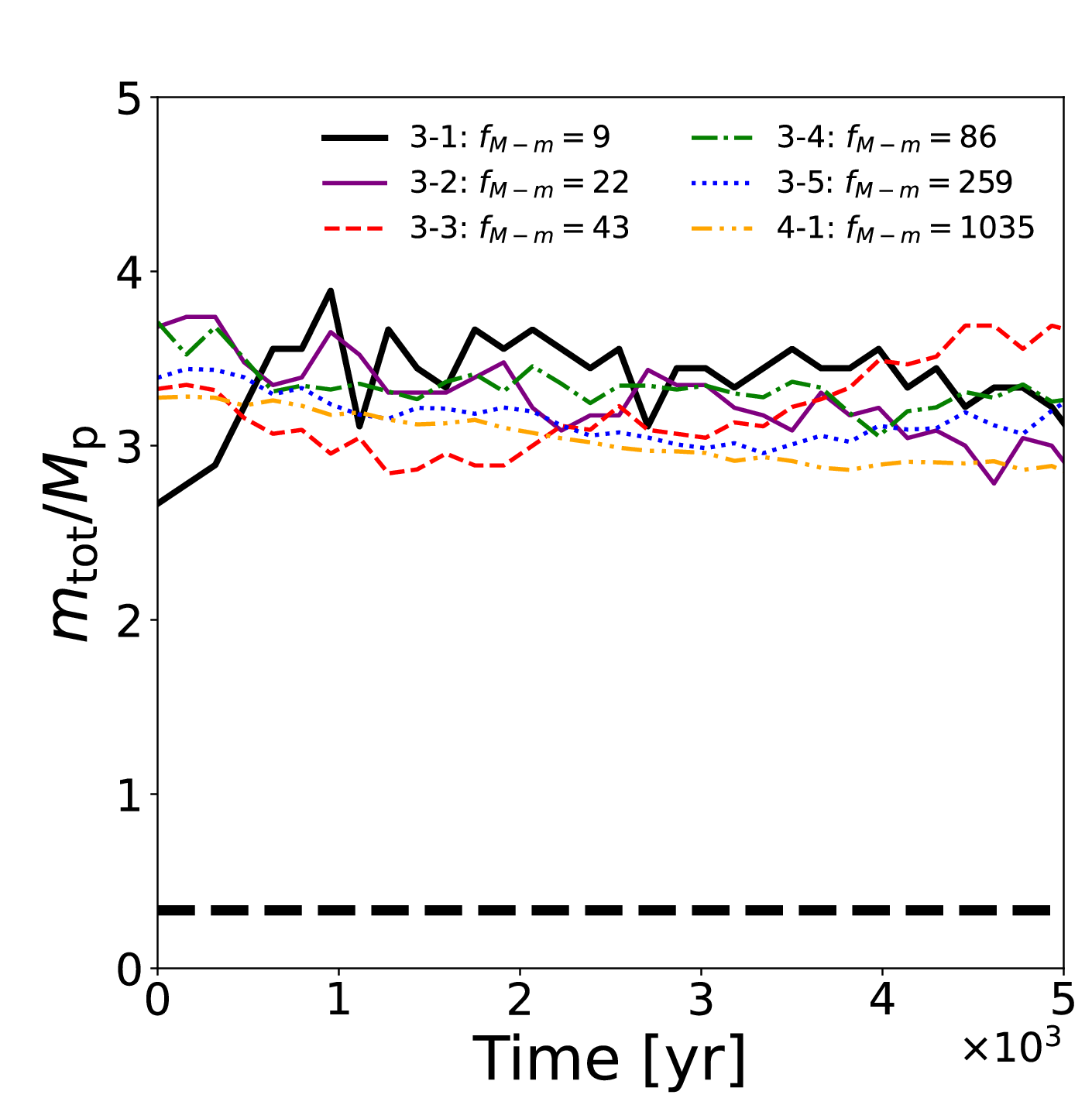}
 \caption{The time evolution of the total planetesimal mass within 5 $r_{\mathrm{Hill}}$ outside the planet normalized by the mass of the planet in models 3-1 to 3-5 and 4-1. The thick solid line (black), thin solid line (purple), dashed line (red), dash-dotted line (green), dotted line (blue), and double-dotted dashed line (orange) represent the results from models 3-1, 3-2, 3-3, 3-4, 3-5, and 4-1, respectively. The thick-dashed black line represents $m_{\mathrm{tot}}/M_{\mathrm{p}}=1/3$, which is the critical value for criterion 1.\\{Alt text: The line graph, with the vertical and horizontal axes representing the total planetesimal mass within 5 $r_{\mathrm{Hill}}$ outside the planet normalized by the mass of the planet and time in thousands of years.}} 
 \label{fig:criteria_1}
\end{figure}
This criterion states that for monotonous PDM to occur, the planet's mass $M_{\mathrm{p}}$ must be within a factor of 3 of the total mass of planetesimals $m_{\mathrm{tot}}$, which are within 5 $r_{\mathrm{Hill}}$ outside the planet. To check whether the mass ratio criterion is satisfied or not, we plot the total mass of planetesimals in front of the planet which exhibits outward PDM in figure \ref{fig:criteria_1}. The thick-dashed black line represents $m_{\mathrm{tot}}/M_{\mathrm{p}}=1/3$, which is the critical value for this criterion. Figure \ref{fig:criteria_1} indicates that this criterion is met in all models. In other words, we have not tested this criterion.

\subsubsection{The mass resolution criterion}\label{discussion:mass_resolution_criterion}
\begin{figure}[hbtp]
 \includegraphics[width= 8cm]{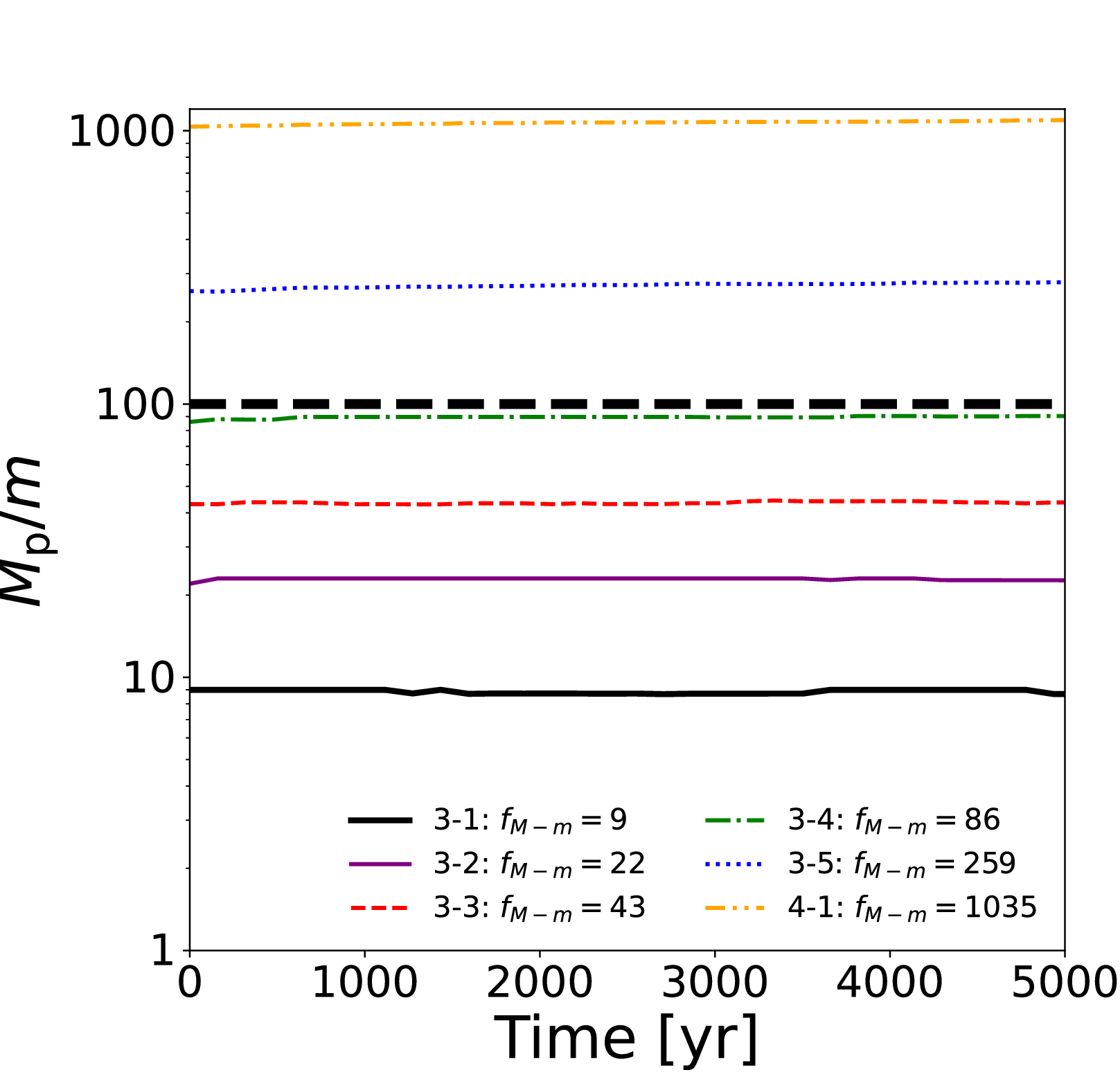}
 \caption{The time evolution of the planet mass normalized by the averaged planetesimal mass within 5 $r_{\mathrm{Hill}}$ outside the planet in models 3-1 to 3-5 and 4-1. The thick solid line (black), thin solid line (purple), dashed line (red), dash-dotted line (green), dotted line (blue), and double-dotted dashed line (orange) represent the results from models 3-1, 3-2, 3-3, 3-4, 3-5, and 4-1, respectively. The thick-dashed black line represents $M_{\mathrm{p}}/m=100$, which is the critical value for criterion 2.\\{Alt text: The line graph, with the vertical and horizontal axes representing the planet mass normalized by the averaged planetesimal mass within 5 $r_{\mathrm{Hill}}$ outside the planet and time in thousands of years.}} 
 \label{fig:criteria_2}
\end{figure}
This criterion tells us that the mass ratio between the planet and the planetesimal should be larger than 100 for the monotonous PDM to occur. To check this condition, we plot the time evolution of the planet mass normalized by the averaged planetesimal mass within 5 $r_{\mathrm{Hill}}$ outside the planet in figure \ref{fig:criteria_2}. According to the figure, only models 3-5 and 4-1 satisfy this criterion. However, we can see that planets in other models also show general trends of migrating outward in figure \ref{fig:criteria_0}. This result indicates that both inward and outward migrations of planets can occur at a lower mass ratio than was previously suggested by \citet{2014Icar..232..118M} and \citet{2016ApJ...819...30K}. 

As we have briefly discussed in section \ref{results:mass_resolution}, \citet{2014Icar..232..118M} used the monotonicity of the migration, -in other words, the condition that a flip in direction does not occur during the simulation- as a criterion for the effectiveness of PDM. What we found is that flip probability is proportional to the inverse of the mass ratio and there is no single critical mass ratio below/above which the flipping occurs or does not occur. We conclude that this mass resolution criterion should be replaced with the new understanding that PDM is always active but flipping is more frequent for smaller mass ratios. This new understanding implies that protoplanets or even planetesimals slightly more massive than others (which naturally form through the runway growth) can radially migrate much more easily than previously assumed. We will study this effect in forthcoming papers.

\subsubsection{The disk eccentricity criterion}\label{discussion:eccentricity_criterion}
\begin{figure}[hbtp]
 \includegraphics[width= 8cm]{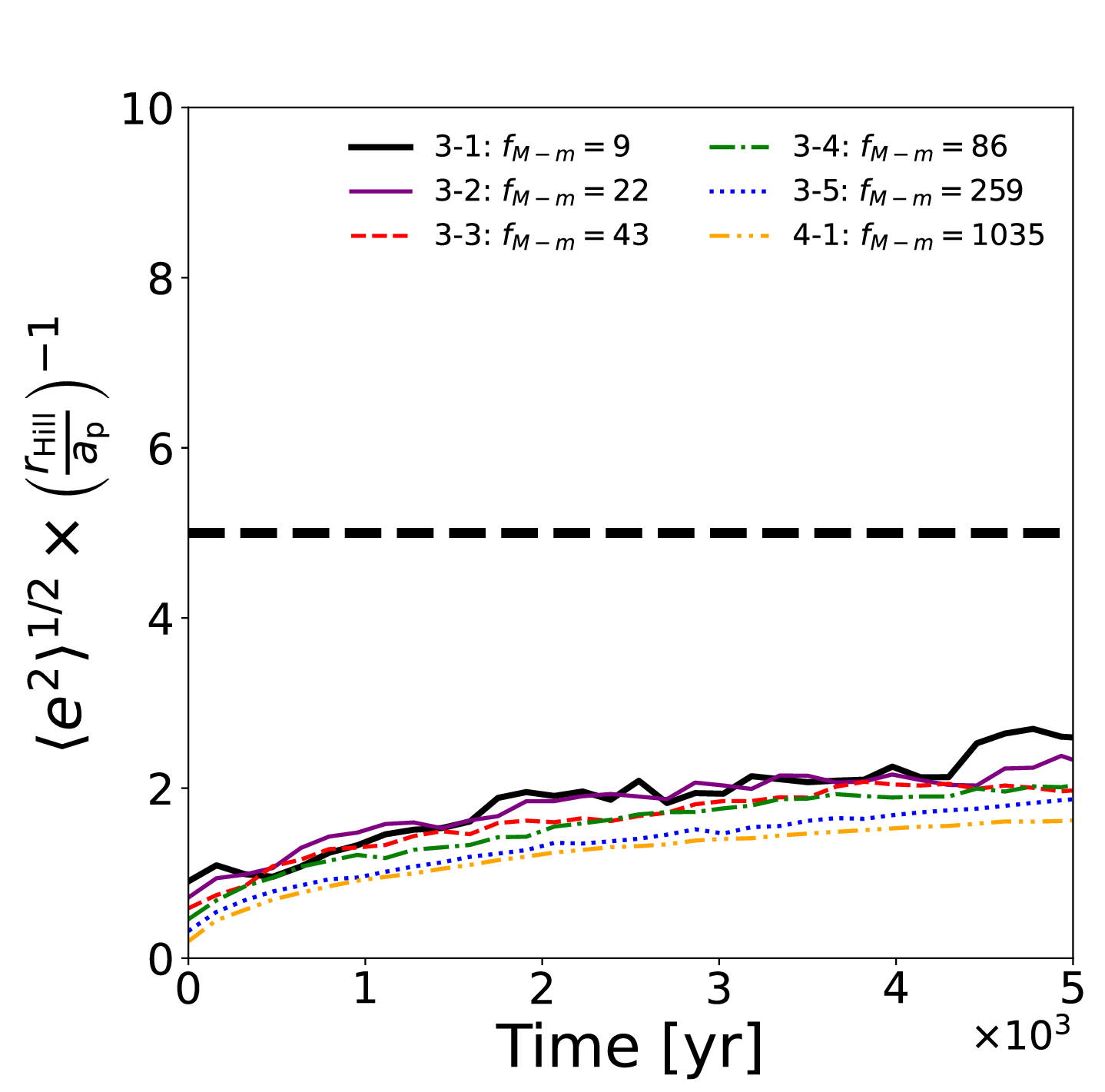}
 \caption{The time evolution of the RMS eccentricity of planetesimals within 5 $r_{\mathrm{Hill}}$ outside the planet normalized by the reduced Hill in models 3-1 to 3-5 and 4-1. The thick solid line (black), thin solid line (purple), dashed line (red), dash-dotted line (green), dotted line (blue), and double-dotted dashed line (orange) represent the results from models 3-2, 3-3, 3-4, 3-5, and 4-1, respectively. The thick-dashed black line represents $\langle e^2 \rangle ^{1/2}\times \left(r_{\mathrm{Hill}}/{a_{\mathrm{p}}}\right)^{-1}=5$, which is the critical value for criterion 3.\\{Alt text: The line graph, with the vertical and horizontal axes representing the RMS eccentricity of planetesimals within 5 $r_{\mathrm{Hill}}$ outside the planet normalized by the reduced Hill and time in thousands of years.}} 
 \label{fig:criteria_3}
\end{figure}
This criterion states that the RMS eccentricity of planetesimals has to be less than $5 \times$ the reduced Hill radius to maintain the constant migration rate. To see this condition, we plot the time evolution of the RMS eccentricity of planetesimals within 5 $r_{\mathrm{Hill}}$ outside the planet normalized by the reduced Hill in models 3-1 to 3-5 and 4-1 in figure \ref{fig:criteria_3}. We can see that the planets in all models meet this criterion. In other words, we have not tested this criterion.

\begin{figure*}[hbtp]
 \begin{center}
    \includegraphics[width= 16cm]{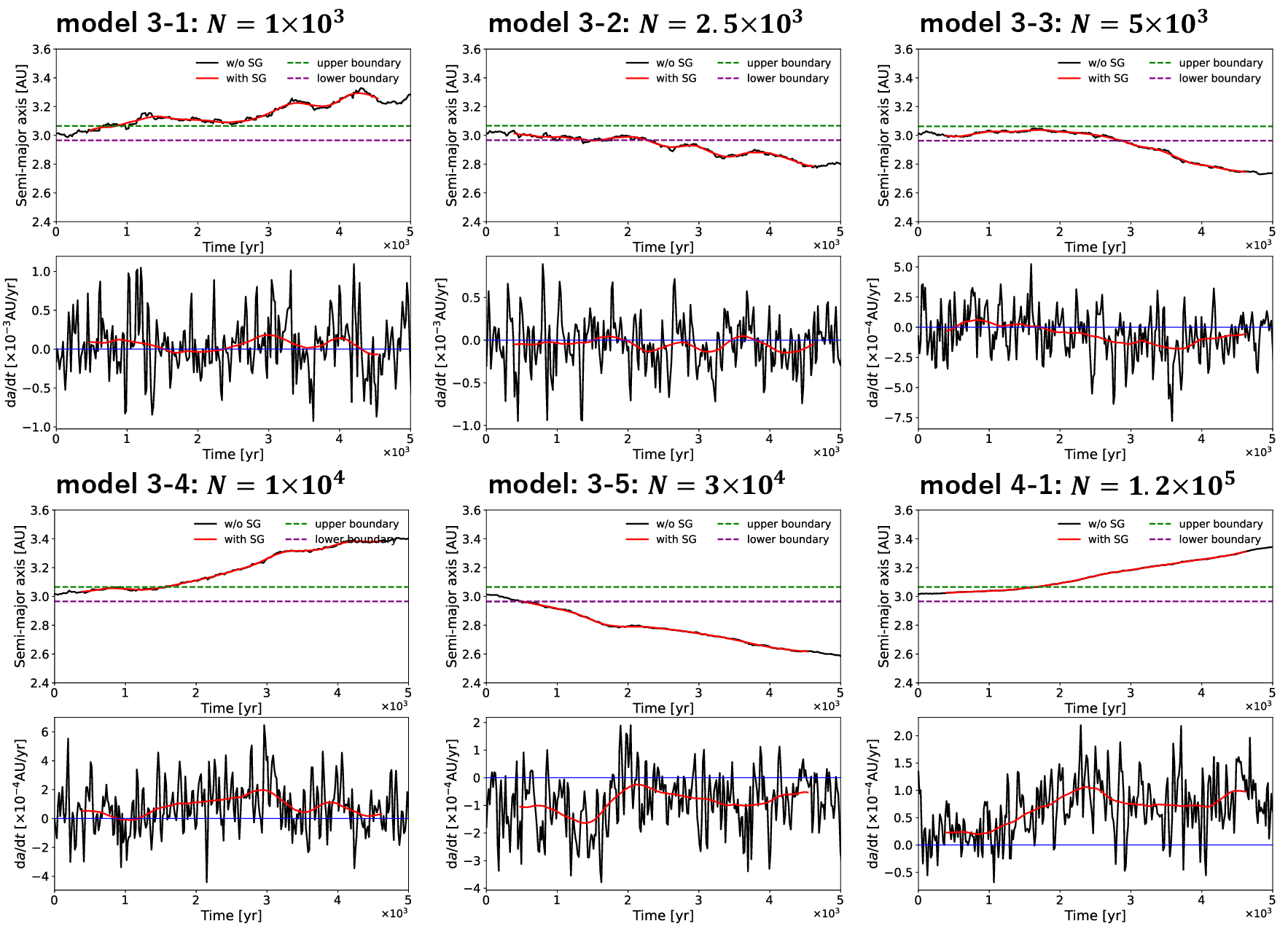}
 \end{center}
 \caption{The time evolution of the planetary semi-major axis (upper panel) and its time derivative (lower panel) for one run each from models 3-1 to 4-1 (from upper left to lower right). The time derivative of the planetary semi-major axis was calculated by finite differences with the sampling rate corresponding to the snapshot interval of our simulations (i.e. $15.9$ years). The red lines represent smoothed planetary semi-major axes and their time derivatives using the Savitzky-Golay (SG) method \citep{1964AnaCh..36.1627S}. The blue lines in the lower panels represent the boundary where the time derivative of the planetary semi-major axis becomes zero. Thus, the points where the blue and red lines intersect mark the moments when the direction of the planet's migration flips. The green and the purple dashed lines represent the semi-major axis $\pm$0.05 AU from the initial planetary semi-major axis, and we define these lines as the upper/lower boundary where PDM begins.\\{Alt text: Line graphs, with 6 subplots from models 3-1 through 3-5 and 4-1. Each subplot has two panels with the vertical and horizontal axes representing the semi-major axis in AU (upper panel) or the time derivative of the semi-major axis in $10^{-4}$ times AU per year (lower panel) and time in thousands of years.}} 
 \label{fig:SI4}
\end{figure*}
\subsection{The effect of two-body effect of planetesimals on PDM}\label{discussion:perturbation}
In sections \ref{results:mass_resolution} and \ref{discussion:mass_resolution_criterion}, we discussed the effect of mass ratio. In this section, we discuss this effect in more detail.

Figure \ref{fig:SI4} shows the time evolution of the planetary semi-major axis (upper panel) and its time derivative (lower panel) for six runs, each selected from models 3-1 to 3-5 and 4-1. The figure's red lines represent smoothed planetary semi-major axes and their time derivatives using the Savitzky-Golay (SG) method \citep{1964AnaCh..36.1627S}. We used an SG method with a second-degree polynomial for data points spanning $4\times10^2$ years before and after each point to smooth the data. A timescale of $4\times10^2$ years is selected due to the synodic periods associated with closed encounters between planetesimals and a planet. Specifically, this timescale is related to encounters between a $0.5$ Earth-mass planet in a circular orbit near 3 AU and planetesimals located at a distance of $1~r_{\mathrm{Hill}}$ from the planet, with an approximate duration of $4\times10^2$ years. 

Figure \ref{fig:SI4} shows that runs with a larger number of particles exhibit fewer flips in planetary migration. This is consistent with the results shown in figure \ref{fig:flipping}, where the flip count of planetary migration direction is proportional to $1/N$. This $1/N$ dependence of planetary migration flips can be explained by two-body relaxation between a planet and planetesimals. Specifically, the variance in velocity changes due to two-body relaxation between a planet and planetesimals is proportional to $1/N$ in the limit where the velocity of a planet becomes zero \citep{2008gady.book.....B}.

Indeed, the increase in velocity dispersion of planets caused by two-body relaxation between a planet and planetesimals slows down the monotonic planetary migration driven by PDM (see figures \ref{fig:trial3_1} and \ref{fig:fiducial}). However, the overall effect of PDM does exist even for a small mass ratio, resulting in efficient radial migration as evident from figure \ref{fig:trial3_1}.

\section{Summary and conclusion}\label{summary_conclusion}
We conducted 570 self-consistent $N$-body simulations of PDM, in which gravitational interactions among planetesimals, the gas drag, and Type-I migration are all taken into account. In paper I, we followed the evolution of a single protoplanet embedded in the radially smoothed planetesimal disk and investigated the effect of PDM on planetary migration. Our main findings are summarized as follows:
\begin{enumerate}
    \item We found that planets can overcome the drift toward the central star caused by Type-I migration, with some planets migrating outward through PDM (figures \ref{fig:model1} to \ref{fig:trial3_1} and \ref{fig:trial4_1}). 
    \item Planets migration through PDM occurs for the mass ratio between protoplanets and planetesimals less than 10. This value is  much
      smaller than those previously proposed  \citep{2014Icar..232..118M,2016ApJ...819...30K} (figures \ref{fig:trial3_1} and \ref{fig:fiducial}).
    \item The degree of monotonicity in PDM is proportional to the number of particles $N$, and this can be explained by the two-body relaxation between planetesimals and planets (figure \ref{fig:flipping}). 
\end{enumerate}

Our results indicate that even during the runaway growth
\citep{1989Icar...77..330W,1993Icar..106..210I,1996Icar..123..180K}, planetary embryos can actively migrate within the protoplanetary disk through PDM. Furthermore, the occurrence of outward PDM in our simulations suggests that the transfer of solid components from the inner to the outer regions of the disk by PDM could potentially enhance the formation of cores in the outer region. Additionally, the fact that outward PDM can be very large implies that the cores of outer planets may initially form in the inner regions of the disk and then migrate outward.

In this paper, we featured an idealized simulation, assuming the existence of a single protoplanet near 3 AU as the initial condition. While this setup is suitable for examining the statistical trends of planetary migration through PDM, it poses challenges in comprehensively investigating the impact of PDM on the general planetary formation process, particularly given the oligarchic growth of multiple protoplanets within the disk \citep{1998Icar..131..171K}. Therefore, in subsequent papers starting with paper II, we explore the effects of PDM on the planet formation process from planetesimals, without assuming the presence of protoplanets in the initial conditions. Specifically, in paper II,  we perform long-term $N$-body simulations on a timescale of $\sim10^6$ years over a large planetesimal disk ($\sim20$ AU), which includes the ice giant formation region. Thus, in paper II, we plan to study not only the effect of PDM on the planetary formation process but also the effect of the disk evolution due to viscous stirring and dynamical friction on PDM.



\begin{ack}
We would like to thank the referee Hanno Rein for providing useful feedback that improved our paper. This work was supported by MEXT as the ``Program for Promoting Researches on the Supercomputer Fugaku" (Structure and Evolution of the Universe Unraveled by Fusion of Simulation and AI; Grant Number JPMXP1020230406), and by JST SPRING, Grant Number JPMJSP2148. Computational resources were provided by the RIKEN Center for Computational Science through the use of the supercomputer Fugaku (Project IDs: hp230204 and hp240094). Test simulations in this paper were also carried out on a Cray XC50 system at the Centre for Computational Astrophysics (CfCA) of the National Astronomical Observatory of Japan (NAOJ).
\end{ack}

\section*{Funding}
This work was supported by MEXT as the ``Program for Promoting Researches on the Supercomputer Fugaku" (Structure and Evolution of the Universe Unraveled by Fusion of Simulation and AI; Grant Number JPMXP1020230406), and by JST SPRING, Grant Number JPMJSP2148.

\section*{Data availability} 
The data that support the findings of this study are available from the corresponding author on request.

\appendix 
\section{The effect of gas-to-dust ratio} \label{appendix:gas-to-dust_ratio}
\begin{figure*}[hbtp]
 \begin{center}
    \includegraphics[width= 16cm]{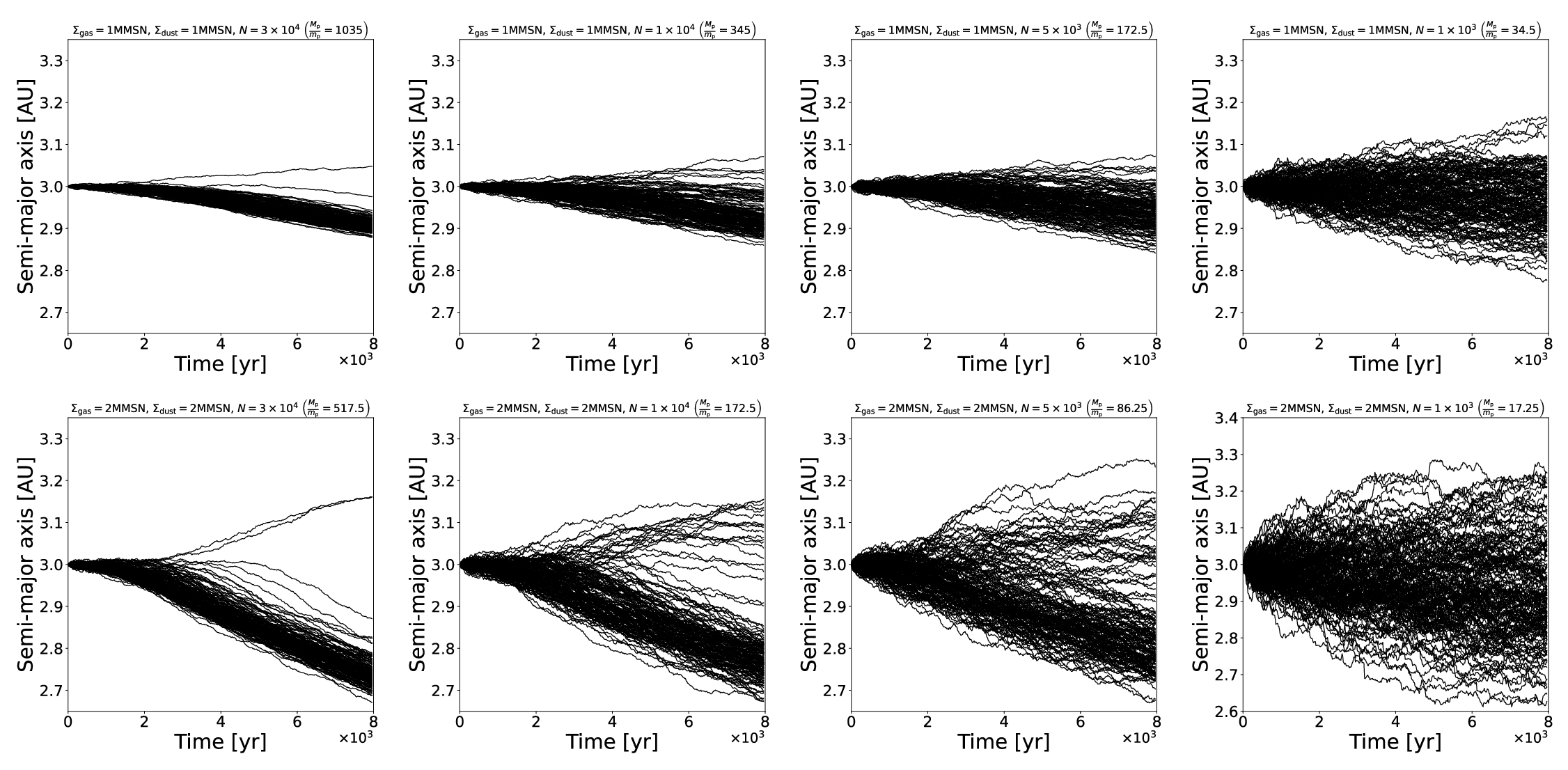}
 \end{center}
 \caption{The time evolution of the semi-major axes of protoplanets embedded in the planetesimal disk. Each panel includes 150 runs with different initial seeds with the same number of particles, totaling 1,200 runs. The initial mass of protoplanets is set to $0.5M_{\oplus}$ in all runs. As shown in the titles of each panel, the upper four panels represent models with $\Sigma_{\mathrm{dust}}$ and $\Sigma_{\mathrm{gas}}$ both set to $1\times$MMSN, while the lower four panels represent models with $\Sigma_{\mathrm{dust}}$ and $\Sigma_{\mathrm{gas}}$ both set to $2\times$MMSN. Additionally, from left to right, the panels show simulation results with the number of particles $N=3\times10^4,~1\times10^4,~5\times10^3$, and $1\times10^3$.\\{Alt text: Line graphs, with 8 subplots. Each subplot's vertical and horizontal axes represent the semi-major axis in AU and time in thousands of years.}} 
 \label{fig:appendix1}
\end{figure*}
The gas-to-dust ratio strongly influences the competition between PDM and Type-I migration. The ratio of Type-I migration speed $v_{\mathrm{Type-I}}$ \citep{2002ApJ...565.1257T} to PDM speed $v_{\mathrm{PDM}}$ (equation (\ref{eq:fiducial})) is given by:
\begin{equation}
\frac{v_{\mathrm{Type-I}}}{v_{\mathrm{PDM}}}\sim\left(\frac{\Sigma_{\mathrm{gas}}}{\Sigma_{\mathrm{dust}}}\right)\left(\frac{M_{\mathrm{p}}} {M_{\mathrm{sun}}}\right)h^{-2}.
\end{equation} 
Thus, whether PDM is dominant over Type-I migration for a given planetary mass is determined in proportion to the gas-to-dust ratio. Here, we show the results of additional simulations using a disk model with a gas-to-dust ratio based on the MMSN to further investigate the competition between Type-I migration and PDM. 

In additional simulations, we employed two disk models with surface densities [$\Sigma_{\mathrm{dust}} = \Sigma_{\mathrm{dust, MMSN}}$ and $\Sigma_{\mathrm{gas}} = \Sigma_{\mathrm{gas, MMSN}}$: 1$\times$MMSN model], as well as [$\Sigma_{\mathrm{dust}} = 2\times\Sigma_{\mathrm{dust, MMSN}}$ and $\Sigma_{\mathrm{gas}} = 2\times\Sigma_{\mathrm{gas, MMSN}}$: 2$\times$MMSN model]. We conducted extra 150 runs for each number of particles, \(N = 1 \times 10^3\), \(5 \times 10^3\), \(1 \times 10^4\), and \(3 \times 10^4\), for both the \(1 \times \mathrm{MMSN}\) and \(2 \times \mathrm{MMSN}\) models, totaling 1,200 runs. In each simulation, a single protoplanet with a mass of 0.5 $M_{\oplus}$ is initially embedded within the planetesimal disk at 3 AU.

Figure \ref{fig:appendix1} shows the time evolution of semi-major axes of protoplanets. Even when accounting for Type-I migration with a gas-to-dust ratio based on the MMSN, protoplanets still exhibit outward migration due to PDM. This result is consistent with the result presented in section \ref{results}. Our additional simulations and results presented in section \ref{results} indicate that outward PDM can universally occur even with factor-level variations in the gas-to-dust ratio, as indicated by observational studies (e.g., MAPS, \cite{PPVII14}).

\bibliographystyle{aasjournal}
\bibliography{reference}
\end{document}